\documentclass[aps,prresearch,reprint,floatfix,longbibliography,superscriptaddress]{revtex4-2}

\usepackage[utf8]{inputenc}
\usepackage[tbtags]{amsmath}
\usepackage{amssymb}
\usepackage{bm}
\usepackage{bbm}
\usepackage{dsfont}
\usepackage{graphicx}
\usepackage{xcolor}
\usepackage[normalem]{ulem}
\usepackage{physics}
\usepackage{gensymb}
\usepackage{tabularx}

\usepackage{hyperref}
\hypersetup{colorlinks,allcolors=black}
\usepackage{cleveref}

\usepackage{soul} % highlighting via \hl{...}; strikeout \sout{...} comes from ulem

\hypersetup{colorlinks=true,linkcolor=blue,citecolor=blue,urlcolor=blue}

\begin{document}

\title{Quantum limit cycles with continuous symmetries from coherent parametric driving: exact solutions and many-body extensions}

% \title{Dissipative quantum limit cycless using parametric driving}

\author{Sihan Chen}
\email[]{sihanc@uchicago.edu}
\affiliation{Department of Physics, University of Chicago, Chicago, Illinois 60637, USA}

\author{Aashish A. Clerk}
\email[]{aaclerk@uchicago.edu}
\affiliation{Pritzker School of Molecular Engineering, University of Chicago, Chicago, Illinois 60637, USA}

\date{April 5, 2026}

\begin{abstract}
There is widespread interest in many-body quantum systems that exhibit limit-cycle or time-crystalline behaviour.  An ideal quantum limit cycle would be realized using fully coherent driving (to minimize noise) and also have a continuous internal symmetry (to ensure generation of monochromatic radiation).  While these two requirements may seem incompatible, we introduce in this work a large class of multi-mode bosonic limit cycle models based on coherent parametric driving which possess an $O(N)$ continuous symmetry.  Surprisingly, the full quantum dissipative steady state of these models can be found exactly.  They exhibit rich physics, including steady state entanglement, reduced phase diffusion and the possibility of realizing quantum limit tori.  The basic mechanism we identify provides a unified way to understand how coherent parametric driving can yield symmetry-enriched limit cycles, and also helps us understand related models where the relevant symmetries are weakly broken.  The models we study are compatible with a range of different experimental platforms, including quantum optical setups and superconducting quantum circuits.   
\end{abstract}

\maketitle

\section{Introduction}

There has been a resurgence of interest in driven-dissipative quantum systems which are analogs of 
classical limit cycle oscillators,
see e.g.~Refs.~\cite{Bruder_PRL_2018, Buca_2022_SciPost, Clerk_PRX_2014, Haque_PRL_2025, Diehl_PRX_2024}. 
 Such systems exhibit long-lived, dynamically stable oscillatory states, and are paradigmatic examples of the breaking of a continuous time-translation symmetry in a non-equilibrium setting, and hence are open system analogues of closed-system time-crystals \cite{Zalatel_RMP_2023, else_2020_review, Sacha_2018_review}. Unlike discrete time crystals characterized by an emergent subharmonic frequency, here the continuous time-translation symmetry is spontaneously broken as arbitrary initial conditions flow to a stable limit-cycle attractor.This attractor requires a balance between external drive and dissipation, making it fundamentally distinct from oscillation arising from generic time-independent Hamiltonian systems \cite{Andronov1966, Solanki2026}.
They are interesting both from a fundamental perspective (e.g.~as a unique kind of non-equilibrium broken-symmetry phase), and for practical applications ranging from sources of coherent and even non-classical light \cite{Clerk_PRX_2014}, to novel platforms for quantum metrology.

\begin{figure}[t]
    \centering
    \includegraphics[width=\linewidth]{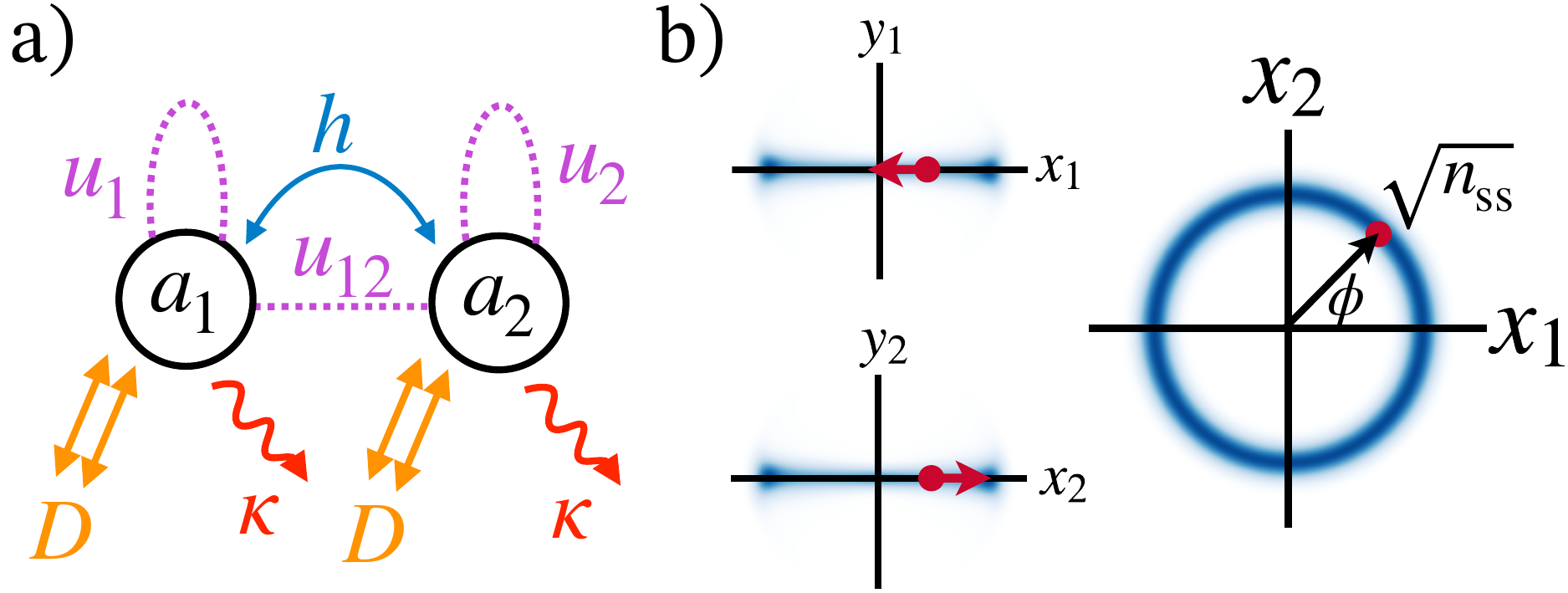}
    \caption{a) Minimal two-mode realization of our coherently-driven many-body limit cycle setup: two bosonic modes with self-Kerr $u_1, u_2$ and cross-Kerr $u_{12}$ nonlinearity tuned to $u_1 = u_2 = u_{12}$, coupled by an imaginary hopping term $ih$, subject to onsite parametric driving $D$, single photon loss $\kappa$. b) Steady-state phase space profile: the single-mode Wigner function $W(x_i, y_i)$ and the joint quadrature probability distribution $P(x_1, x_2) = \mel{x_1, x_2}{\hat{\rho}}{x_1, x_2}$. The red dots and arrow indicate the semiclassical trajectory of the limit cycles. The state forms a uniform ring in the $(x_1, x_2)$ plane with radius $\sqrt{n_{\rm ss}}$, and angle $\phi$ labels the limit cycles phase. The behavior depicted here generalizes to an $N$-mode system, with the relevant continuous symmetry being $O(N$). Parameters for (b): $u=0.02, D=\kappa=1$} 
    \label{fig1}
\end{figure}

Among the many examples of quantum limit cycles, a special privileged role is played by those that possess an internal continuous symmetry (e.g.~a $U(1)$ symmetry).  The continuous symmetry directly defines natural phase degrees of freedom, phases that wind at a constant rate in the limit cycle state.  Such systems produce monochromatic light, and are relevant to applications in the same way as a conventional laser or microwave oscillator.  Continuous symmetry also greatly simplifies the description of dynamics, as well as the synchronization physics that results when several individual limit cycle systems are coupled.  The standard route to realizing a $U(1)$ or $O(N)$-symmetric quantum limit cycles is to use incoherent driving, i.e.,~driving that does not have any natural phase reference, and that is realized by coupling to a dissipative environment.  Classic examples include bosonic systems like a standard laser with an incoherent gain medium \cite{scully_zubairy_1997} and the quantum van der Pol oscillator \cite{Cross_PRR_2021, Bruder_PRL_2016, Dykman_Krivoglaz_1984}.  They also include spin-based limit cycles like the superradiant laser \cite{Thompson_Nature_2012}.

We now see what seems to be a fundamental trade-off:  if we want a dissipative quantum system with a continuous symmetry that also has limit cycle physics, we must necessarily introduce extra dissipation into the system, via the environment that creates the phase-agnostic incoherent pumping.  If the goal is to minimize noise and enhance quantum effects, this tradeoff is especially unfortunate.  We note that the extra quantum noise associated with incoherent pumping contributes half the phase noise that sets the celebrated Schawlow-Townes limit on the linewidth of a laser \cite{schawlow_infrared_1958, Wiseman_laser_SQL}.

In this paper, we show that the apparent incompatibility between continuous symmetry and low dissipation is not fundamental.  We present a surprisingly simple route for generating an exactly-solvable family of quantum limit cycle systems (both few and many-body) that have $O(N)$ continuous symmetry, but that do not involve the extra dissipation and noise associated with an incoherent drive.  The ingredients we exploit are ubiquitous across many systems: $N$ bosonic modes with Kerr-type nonlinearities, and coherent parametric (or two-photon) drives.  
The key observation is that while parametric drives break the $U(1)$ phase symmetry of each individual mode, a collection of $N$ identically driven modes with balanced nonlinearities can still retain a weak $O(N)$ symmetry in mode space.
The minimal realization of this physics involves two parametrically-driven modes whose combined dynamics retains a weak $U(1)$ symmetry (see Fig. \ref{fig1}a). Through a semiclassical analysis, we show indeed that this setup exhibits phase diffusion noise that is suppressed by a factor of two compared to the standard Schawlow–Townes limit \cite{schawlow_infrared_1958, Wiseman_laser_SQL}.

Despite its complexity (nonlinearity, dissipation, driving), we are also able to study properties of the full quantum description of our system without approximations:  it admits (for any number of modes) an exact analytical solution for its non-equilibrium steady state. 
Unlike simpler quantum limit cycles, our setup necessarily involves multiple degrees of freedom, allowing one to meaningfully ask questions about steady-state entanglement (see also \cite{PhysRevA.108.062216, Conti_PRA_2025}).  For $N \geq 4$, we also show that our setup can realize many-body quantum limit tori systems, which are higher-dimensional generalizations of limit cycles (something studied recently using incoherent driving \cite{Nowoczyn2025Universal}).

Even in the minimal $N=2$ case, our exact solution reveals properties with no counterpart in prior analyses of parametrically-driven setups, including non-Gaussian intermode entanglement that saturates at a finite value deep in the semiclassical large-photon-number regime. This minimal model also serves as a unifying framework for understanding several models studied independently in the literature: $U(1)$ symmetry breaking in non-degenerate parametric oscillators \cite{Graham_PRA_1970, Reid_Drummond_PRA_1989, Roukes_PRL_2012, Iacopo_PRA_2007}, finite-amplitude limit cycle phases in coherently driven coupled parametric oscillators \cite{bello_persistent_2019, Cosme_PRA_2024}  (which can be understood as symmetry-broken perturbations of our model), and even realizations of cold-atom supersolidity based on the emergence of a $U(1)$ symmetry through the symmetric coupling of two distinct $\mathbb{Z}_2$ symmetries \cite{Leonard_Nature_2017}.

\section{Parametrically driven quantum limit cycle with $O(N)$ symmetry}
The key ingredients required to realize an $O(N)$ symmetric quantum limit cycle are $N$ nonlinear resonant bosonic modes, each subject to a resonant parametric drive of the same amplitude $D$. In the rotating frame set by the parametric drives, the Hamiltonian is
\begin{align} \label{n_mode_model}
    \hat{H} &= u \left( \sum_{i=1}^{N} \hat{n}_i \right)^2 - D \sum_{i=1}^{N} (\hat{a}_i^\dagger \hat{a}_i^\dagger + \text{h.c.}) - i h \sum_{i,j=1}^N K_{ij} \hat{a}^\dagger_i \hat{a}_j 
\end{align}
with photon number operator $\hat{n}_i = \hat{a}^\dagger_i \hat{a}_i$, and $K$ a real antisymmetric matrix. The coefficient $u$ describes a Kerr nonlinearity that depends only on the total photon number, and $h$ characterizes the strength of the antisymmetric hopping term. 

The full dynamics of the system also includes independent single-photon loss at rate $\kappa$ on each mode, leading to the Lindblad master equation \cite{Lindblad1976, Gorini:1975nb}
\begin{align}
    \label{eq:Lindblad}
    \partial_t \hat{\rho} = -i[\hat{H}, \hat{\rho}] + \kappa\sum_{i=1}^N \mathcal{D}[\hat{a}_i] \hat{\rho} 
\end{align}
with $\mathcal{D}[\hat{X}]\hat{\rho} = \hat{X} \hat{\rho} \hat{X}^\dagger - (1/2)\{\hat{X}^\dagger \hat{X}, \hat{\rho} \}$.

We now ask about the symmetries of our model.  At first glance, a parametric (i.e.,~two photon) drive has a well defined phase, and thus only leaves one with a $\mathbb{Z}_2$ symmetry.  There is no invariance under arbitrary gauge transformations of each mode, only invariance under a phase shift of $\pi$, i.e.,~$\hat{a}_j \rightarrow (-1) \hat{a}_j$.  For a single mode, this would indeed be the only symmetry.  However, in our $N$-mode system with symmetric drive amplitudes and balanced interaction parameters, we have a much larger symmetry.  Although the parametric drive indeed picks out a preferred phase (up to a sign) for each of the modes, both the driving term and the nonlinear interaction are invariant under orthogonal rotations in mode space. Explicitly, consider first the case $h=0$. In this limit, the Hamiltonian and the full Lindbladian are invariant under transformations $\hat{a}_i \rightarrow \hat{a}'_i = O_{ij}\hat{a}_j $, where $O \in O(N)$. This corresponds to a weak symmetry of the Lindbladian \cite{Liang_PRA_2014, Buca_Prosen_NJP_2012} because the full Lindbladian remains invariant even though the individual jump operators do not. The corresponding symmetry generators are
\begin{align}
    \hat{Q}_{ij} = i(\hat{a}_i^\dagger \hat{a}_j - \hat{a}^\dagger_j \hat{a}_i)
\end{align}

The role of the antisymmetric coupling term $\propto h$ is now apparent: it is proportional to a sum of the $O(N)$ symmetry generators. More generally, one may add generators of any continuous subgroup of $O(N)$, which produce a limit cycle when the resulting trajectory is closed. Given our $O(N)$ symmetry, when the Lindbladian has a unique steady-state (as will always be the case here), the steady-state density matrix must be invariant under $O(N)$ rotations of the modes. Therefore, adding the antisymmetric coupling matrix $K$ reduces the symmetry of the Lindbladian to the stabilizer subgroup of $K$, while leaving the steady-state density matrix unchanged. However, it will affect the long-time dynamics. In the semiclassical description, the dynamics without the antisymmetric coupling will relax onto a continuous family of symmetry-related fixed points. Adding antisymmetric coupling then generates persistent oscillations along this attractor manifold. In the large photon number limit, this motion persists indefinitely and realizes sustained oscillations. 

In the following sections, we show that the $N=2$ case yields the simplest setting where one has $O(2) \cong U(1) \rtimes \mathbb{Z}_2$ symmetry and realizes quantum limit cycle behavior. The analysis can be straightforwardly extended to $N$-mode case. Interestingly, for $N \geq 4$, our model can realize quantum limit tori with quasiperiodic motion, whose frequencies are determined directly by the antisymmetric matrix $K$.

%%%%%%%%%%%%%%%%%%%%%
\section{Minimal two-mode model: parametric driving that retains $U(1)$ symmetry}
We now consider the minimal realization of the more general $O(N)$ model in the case when $N=2$ (see Fig.~\ref{fig1}a). This case is the simplest to analyze, while already capturing the essential mechanism that underlies our larger family of arbitrary $N$ quantum limit cycle systems.  For two modes, the Hamiltonian is 
\begin{equation} \label{two_mode_model}
    \begin{aligned}
        \hat{H} &= u (\hat{a}_1^\dagger \hat{a}_1 + \hat{a}^\dagger_2 \hat{a}_2)^2 - D(\hat{a}_1^\dagger \hat{a}_1^\dagger + \hat{a}_2^\dagger \hat{a}_2^\dagger + \text{h.c.} ) \\
    &- ih(\hat{a}_1^\dagger \hat{a}_2 - \hat{a}_2^\dagger \hat{a}_1).
    \end{aligned}
\end{equation}
For convenience, we take the hopping term to be imaginary, which  simply corresponds to a useful choice of phase reference or gauge for the two modes. Since $O(2) \cong SO(2) \rtimes \mathbb{Z}_2$, the Lindbladian has a continuous weak $SO(2)$ symmetry corresponding to real rotation between the two modes.

The weak $SO(2)$ symmetry of our Lindbladian is equivalent to a weak $U(1)$ symmetry. This is most explicit if we write our Hamiltonian using a rotated mode basis $\hat{b}_1 = (\hat{a}_1 + i \hat{a}_2)/\sqrt{2}, \hat{b}_2 = (\hat{a}_1 - i \hat{a}_2)/\sqrt{2}$:
\begin{align} \label{b_mode_eq}
    \hat{H} = u\hat{N}^2 - 2D\left( \hat{b}_1^\dagger \hat{b}_2^\dagger + \hat{b}_1 \hat{b}_2 \right) - h (\hat{b}_1^\dagger \hat{b}_1 - \hat{b}_2^\dagger \hat{b}_2) ,
\end{align}
with $\hat{N} = \hat{b}_1^\dagger \hat{b}_1 + \hat{b}_2^\dagger \hat{b}_2$. In this basis, the weak $U(1)$ symmetry corresponds to the transformation $\hat{b}_1 \rightarrow e^{i\theta} \hat{b}_1, \hat{b}_2 \rightarrow e^{-i\theta} \hat{b}_2$, corresponding to a rotation of the relative phase between the two modes. Written in this basis, our model is directly connected to the well-known non-degenerate parametric oscillator (NDPO), for which the presence of $U(1)$ symmetry is familiar \cite{Graham_PRA_1970, Reid_Drummond_PRA_1989, Roukes_PRL_2012, Iacopo_PRA_2007}. This connection is conceptually useful, but for our purposes the original $\hat{a}_i$ basis provides the more natural formulation, since it generalizes directly to the $N$-mode case where the full $O(N)$ symmetry is manifest.
We also note that even in the $N=2$ case, our setup differs from more standard NDPO setups where the nonlinearity corresponds to pump depletion and is dissipative in nature; in contrast, we have fully coherent Kerr-type nonlinearities.   
Note that for $N=2$, Eq.~\eqref{b_mode_eq} represents a potentially attractive way to implement our scheme: while Eq.~\eqref{two_mode_model} requires two parametric drives (one on each mode $\hat{a}_j$), Eq.~\eqref{b_mode_eq} only requires a single two-mode squeezing drive.  Further, it does not require any linear beam-splitter coupling between the modes, as the parameter $h$ now just corresponds to the difference between the resonance frequencies of the two modes.   

%%%%%%%%%%%%%%%
\section{Semiclassical analysis and emergence of limit cycles}

\subsection{Semiclassical dynamics}
To obtain a picture of the limit cycle that emerges in our model, we first analyze the semiclassical limit of the master equation. In the limit of large photon number, the dynamics is well-captured by a mean-field ansatz where we replace  $\hat{a}_i(t) \rightarrow \left( a_i(t) \equiv \langle \hat{a}_i(t)\rangle \right)$. The resulting equations of motion are
\begin{align}\label{semi_coherent_eq}
    \partial_t a_i = -i 2u n a_i + i 2D a_i^* - \frac{\kappa}{2}a_i + h \sum_{j=1}^2 J_{ij}a_j,
\end{align}
where $n = |a_1|^2 + |a_2|^2$ and $J = -i\sigma_y$ is a 2 $\times $ 2 matrix. 

At a heuristic level, when the parametric drive $D$ is larger than the loss rate, the vacuum solution $a_i = 0$ becomes unstable, and the system flows to a large amplitude state (whose form is set by balancing the driving against the nonlinearities $u$).  
The instability of the vacuum solution is analogus to a standard parametric oscillator:
the drive $D$ generates effective anti-damping for one quadrature of each mode, giving rise to parametric instability that is eventually stabilized by the Kerr nonlinearity. 

Given the connection to parametric instability, it is useful to write the equations in an appropriate quadrature basis.  We introduce two real-valued lab-frame quadrature vectors $\vb{x}=(x_1, x_2), \vb{y}=(y_1, y_2)$ and write $\vb{a} = e^{i\theta}(\vb{x} + i \vb{y})$.   We pick $\theta = \frac{1}{2}\arcsin(\kappa/4D)$, as this makes the resulting $x$-quadrature dynamics especially simple:  the $\kappa$-induced damping of the quadrature is now exactly canceled by the effective anti-damping generated by the coherent drive. With this choice of quadrature basis, the parametric instability causes the $x_j$ quadratures to saturate to a large finite value, while the $y_j$ quadratures are damped to zero. We further define rotating-frame quadrature amplitudes $\widetilde{\vb{x}}, \widetilde{\vb{y}}$ given by $\widetilde{\vb{x}} = e^{-hJt}\vb{x}$ and $\widetilde{\vb{y}} = e^{-hJt}\vb{y}$, where $e^{-hJt}$ is a matrix exponential in $SO(2)$ describing rotation at angular frequency $h$. In terms of these variables, the semiclassical equation of motion becomes
\begin{align} \label{quadrature_EOM}
    \partial_t \widetilde{x}_i &=  \left(2un + 2D \cos(2\theta) \right) \widetilde{y}_i  \\
    \partial_t \widetilde{y}_i &=  \left(-2un + 2D \cos(2\theta) \right) \widetilde{x}_i - \kappa \widetilde{y}_i
\end{align}
As expected, the linear damping at rate $\kappa$ causes the $\widetilde{y}$ quadratures to relax to zero, i.e.,~$\widetilde{y}_i \rightarrow 0$, while the $\widetilde{x}$ quadratures grow to a non-zero value that is set by the nonlinearity. For $\widetilde{y}_i = 0$ and $ 2un =  2D \cos(2\theta)$, we have $\partial_t \widetilde{x}_i =0$. Since the transformation from $(\widetilde{\vb{x}}, \widetilde{\vb{y}})$ to $(\vb{x}, \vb{y})$ is a rotation, it does not change the steady state amplitude. It follows that the semiclassical attractor manifold is characterized by $y_i = 0$ and a continuous ring in the $x_i$ plane defined by
\begin{align}
    x_1^2 + x_2^2 = \frac{D}{u}\sqrt{1-\kappa^2/16D^2} \equiv n_{\rm ss}.
\end{align}

In the rotating frame, each point on the ring is a valid steady-state solution to the equations of motion; they are all related to each other by the action of the symmetry transformation. Semiclassically, one can interpret selecting a point on the steady-state circle as a spontaneous breaking of the $U(1)$ symmetry. The same structure naturally extends to the general $O(N)$ model, where the attractor manifold becomes generalized spheres $S^{N-1}$. We defer the analysis to Sec. \ref{sec_N_mode}. 

In the rotating frame, the combination of the parametric drive, nonlinearity and loss pins the $x$ quadrature amplitudes to lie on a ring of radius $\sqrt{n_{\rm ss}}$.  If we return to the lab frame, the system will oscillate along this ring at a frequency $h$: we thus have as advertised a limit cycle oscillator in a system where the driving is fully coherent, and there is a continuous $U(1)$ (i.e.,~$SO(2)$) symmetry.  Note that the $\mathbb{Z}_2$ symmetry breaking associated with a single parametric drive still has its vestiges in our limit cycle: each mode only involves excitation of the $x$ quadrature, i.e.,~the phase of each individual mode is fixed modulo $\pi$.  We also note that our limit cycle is inherently delocalized between two modes (i.e.,~the limit cycle oscillations involve energy sloshing between the two localized modes $a_1, a_2$).  This will have interesting consequences when we analyze the full quantum version of our system (namely, the limit cycle will be associated with non-zero entanglement between the modes).

In the standard way, we can study the stability of the limit cycle solution by looking at the linear stability of small deviations.  Linearization around the steady state shows that, as expected, radial fluctuations away from the limit cycle (involving either $x$ or $y$ quadratures) decay to zero 
(see App.~\ref{app_stability}). In contrast, fluctuations along the limit cycle circle in the $(x_1,x_2)$ plane correspond to a Goldstone mode, and have no restoring force.  If we parameterize the limit cycle ring by $\vb{x} = \sqrt{n_{\rm ss}}\hat{\vb{r}}$ with $\hat{\vb{r}} = (\cos \phi, \sin \phi)$, we find that in the original lab frame, the equation of motion for the phase variable $\phi(t)$ is
\begin{align} \label{phase_EOM}
    \partial_t \phi = h + 4 u \sqrt{n_{ss}} \delta y_{\|}, \qquad  \partial_t (\delta y_{\|}) = -\kappa \delta y_{\|}
\end{align}
where $\delta y_{\|}$ is the linearized $y$-quadrature fluctuation tangential to the limit cycle. In the long-time limit, $\delta y_{\|}$ damps away, resulting in persistent oscillation with phase governed by
\begin{align}
    \phi(t) = h t.
\end{align}
Interestingly, only one of the three decaying transverse fluctuation modes couples to the phase variable (i.e.,~only the mode $\delta y_{\|}(t)$). The long-time dynamics of our two-mode limit cycle setup is depicted in Fig.~\ref{fig1}b.    Our semiclassical analysis can be extended to also treat fluctuation effects in the standard manner, we turn to this analysis in Sec. \ref{sec_phase_diffusion}. 

\subsection{Mechanical Analogy}

Despite the simplicity of our semiclassical analysis, the basic intuition underlying the emergence of limit cycles in our setup might remain puzzling.  This is because limit cycles are inherently a dissipative phenomena, and in our discussion above, dissipation seems to play a minimal role.  In this brief subsection, we show that   
additional insight can be provided by viewing
the quadratures $(x_1,x_2)$ as the Cartesian position co-ordinates of a fictitious particle moving in two dimensions.  With this mapping (essentially a reinterpretation) of our dynamics, the parametric drive will correspond to a fully conservative force.  In contrast, the single photon loss $\kappa$ will now play a surprising and non-trivial role: it maps to an effective non-reciprocal interaction, in the form of a so-called ``non-Hookian" spring.  Such non-reciprocal elastic forces have been the subject of much recent interest, in particular in the context of 
odd-elasticity \cite{Scheibner2020}.  Within this mapping, the non-reciprocal non-Hookian spring is able to perform net work on the system, thus stabilizing a self-sustained oscillation state. More broadly, this mapping places our model in a wider class of models exhibiting dissipation-induced instabilities \cite{Marsden_RMP_2007}, where dissipation plays the counterintuitive role of destabilizing the system's dynamics, leading to persistent oscillation. Related ideas have appeared, for example, in the context of atomic Bose-Einstein condensate in an optical cavity \cite{Esslinger_Science_2019}.

To make this mapping explicit, we transform 
Eq.~(\ref{quadrature_EOM}) back to the lab frame in terms of $\vb{x}, \vb{y}$. Near the attractor manifold, we approximate the total photon number $n$ as a slowly varying function of $\vb{x}$ only. This allows us to eliminate $y_i$ and obtain an effective equation of motion involving only $x_i$.  Defining a potential energy:
\begin{equation}
    U(\vb{x}) = \frac{2}{3}u^2 |\vb{x}|^6 - \frac{1}{2}[(2D\cos(2\theta))^2 + h^2]|\vb{x}|^2, 
\end{equation}
the equation of motion for our fictitious particle takes the form
\begin{align}
    \partial_t^2 \vb{x} = -\nabla_{\vb{x}} U(\vb{x}) + \kappa h J \vb{x} + (-\kappa I + 2h J) \partial_t \vb{x}
\end{align}
where $J = -i\sigma_y$ and $I$ is $2 \times 2$ identity matrix. Within this mapping, the parametric drive acts like a negative mass that triggers a spontaneous breaking of $U(1)$ symmetry of the $|\vb{x}|^6$ potential, producing a minimum at
\begin{align}
    |\vb{x}|^2 = \frac{\sqrt{(2un_{\rm ss})^2 + h^2}}{2u}.
\end{align}
Note that this minimum is different from the limit cycle radius, which occurs at radius $|\vb{x}|^2 = n_{\rm ss}$. The imaginary hopping term acts like an effective magnetic field perpendicular to the $(x_1, x_2)$ plane, i.e., they generate an effective Lorentz force. Most interestingly, there is a linear restoring force with an antisymmetric stiffness matrix $[K_{\rm odd}]_{ij} = \kappa h J_{ij}$ (where recall that the matrix $J = -i \sigma_y$). Unlike a spring obeying Hook's law, the forces associated with $K_{\rm odd}$ cannot be written as the gradient of some potential \cite{Scheibner2020}.  As a result, these terms act like an effective energy source that generates a non-reciprocal force $F_i = \kappa h J_{ij}x_j$ along the ring.  This in turn yields persistent oscillatory motion along the symmetry-broken manifold. We thus see that our parametrically driven two-mode system has a direct analogy to a novel two-dimensional mechanical limit cycle based on a ``non-Hookian" spring.  Alternate routes for limit cycles based on such springs were discussed recently in Ref. \cite{Vincenzo_Nature_2025}.

\begin{figure*}[t]
    \centering
    \includegraphics[width=\textwidth]{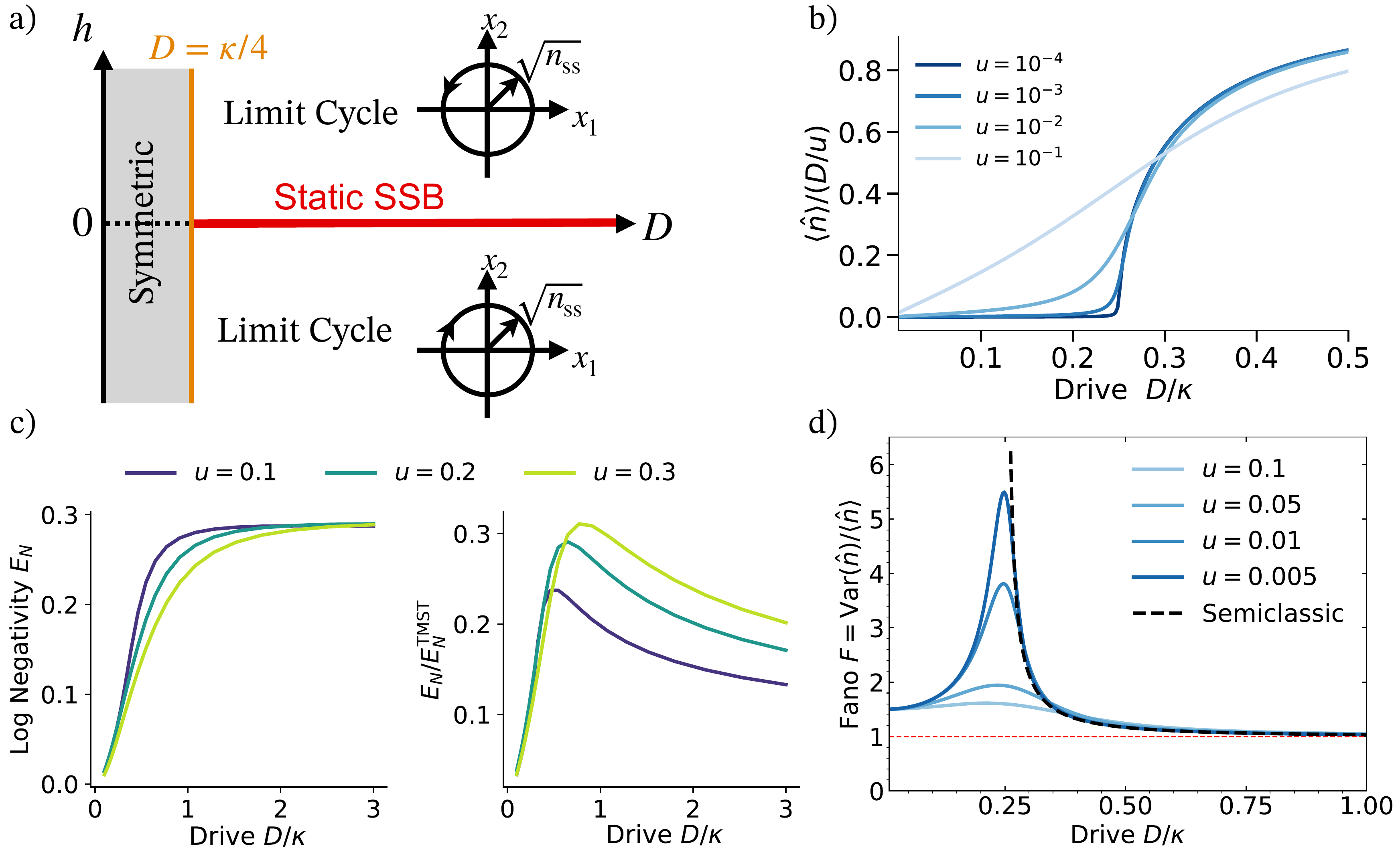}
    \caption{a) Phase diagram of the model for $N=2$. Three phases are identified: limit cycles phase ($h \neq 0, D > \kappa/4$), static phase with spontaneous symmetry breaking (SSB), and a symmetric phase. The orange line marks the transition between the SSB phase and the symmetric phase. b) Order parameter characterizing the transition from SSB to symmetric phase for various nonlinearities $u$, setting $\kappa=1$. The limit $D/u \rightarrow \infty$ corresponds to the thermodynamic limit, and the transition is second order. c) Log-negativity $E_N$ and the ratio of $E_N$ to that of a two-mode squeezed thermal state of the same purity and total photon number. With $u=0.1, D=3$, there are $n = 30$ total photons in the cavity.  d) Fano ratio for various strength of Kerr nonlinearity $u$ at fixed $\kappa=1$ as a function of driving strength $D$.}
    \label{fig2}
\end{figure*}

\section{Exact quantum description of an $O(N)$ symmetric, coherently driven limit cycle}

\subsection{Exact solution and non-equilibrium steady-state phase diagram}

We now turn to the full quantum description of our $N$-mode setup, Eq.~\eqref{eq:Lindblad}.  This dynamics is a priori non-trivial, given the nonlinear interactions, dissipation, and driving.  Surprisingly, despite these complications one can find an exact analytic description of the non-equilibrium steady state (NESS) of  Eq.~\eqref{eq:Lindblad} for arbitrary parameter choices \cite{two_photon_exact}. The exact description of the NESS will both allow us to understand aspects of the system's phase diagram and quantum features of the dynamics such as entanglement.  

The exact NESS $\hat{\rho}_{\rm ss}$ of our $N$-mode model is most conveniently expressed in terms of a $2N$-mode purification. Concretely, we introduce an auxiliary copy of the original systems so that the pure state $\ket{\psi}$ lives in a $2N$-mode Hilbert space consisting of $N$ physical modes, labeled by $L$, and $N$ auxiliary modes, labeled by $R$. The steady-state is then obtained by tracing out the auxiliary copy:
\begin{align}
    \hat{\rho}_{\rm ss} &= \Tr_R \ket{\psi}\bra{\psi}.
\end{align}
One finds \cite{two_photon_exact}:
\begin{align}
    \ket{\psi} &= \sum_{m=0}^\infty \frac{(D/u)^m}{m!(\delta)_m} \left(\frac{1}{2}\sum_{i=1}^N \hat{\alpha}_i^\dagger \hat{\alpha}_i^\dagger \right)^m \ket{0}_L \ket{0}_R.
\end{align}
Here $\ket{0}_L\ket{0}_R$ is the vacuum of all $2N$ modes, $\hat{\alpha}_i = (\hat{a}_{i,L} + \hat{a}_{i,R})/\sqrt{2}$ is the symmetric combination of the left and right modes, and $(\delta)_m = \delta(\delta+1)\cdots(\delta+m-1)$ with $\delta = 1 - i \kappa/4u$, see App. \ref{app_exact}. The steady-state is a condensate of boson pairs in the symmetric mode basis with no excitation in the antisymmetric basis.

In many-body systems, the thermodynamic limit is taken by sending volume $V \rightarrow \infty$. The analogous extensive parameter in our system is the total photon number \cite{ciuti_PRA_2017}, which scales as $n_{\rm ss} \propto D/u$. Without Kerr nonlinearity $u$, our model reduces to a linear degenerate parametric amplifier with stability controlled by $D/\kappa$. At a critical drive strength, the linear system becomes unstable due to parametric amplification, leading to divergent photon number. A weak Kerr nonlinearity cuts off the instability and stabilizes the system at large photon number, but it does not change the underlying instability threshold. This suggests that the appropriate thermodynamic limit $n_{\rm ss} \rightarrow \infty$ is realized by taking $u \rightarrow 0$ while holding the drive $D$ fixed, which we expect to probe the parametric instability transition controlled by $D/\kappa$. This is distinct from setting $u = 0$, as a strictly non-interacting system does not admit a stable steady-state, whereas a non-zero $u \rightarrow 0$ stabilizes a steady-state. An alternative justification of our scaling is given in App. \ref{app_TD_limit}. 

Given the above, we will consider in what follows the rescaled photon number as our order parameter: 
\begin{align}
    \lim_{u \rightarrow 0} \frac{\langle \hat{n} \rangle}{D/u},
\end{align}
where $\hat{n} = \sum_i \hat{a}_i^\dagger \hat{a}_i$. From App.~\ref{app_obs}, we find the exact steady state photon number to be
\begin{align}
    \langle \hat{n} \rangle = \lambda \frac{(N/2)}{|\delta|^2} \frac{{}_1 F_2(N/2+1; \delta+1, \delta^* + 1; \lambda)}{{}_1 F_2(N/2; \delta, \delta^*; \lambda)}
\end{align}
where $\lambda = (D/u)^2$ and ${}_p F_{q}$ is hypergeometric function. Evaluating the $u \rightarrow 0$ limit, we find 
\begin{align}
    \lim_{u \rightarrow 0} \frac{\langle \hat{n} \rangle}{D/u} = 
    \begin{cases}
        0, &\qquad D < \kappa/4 \\
        \text{finite}, &\qquad D > \kappa/4
    \end{cases}
\end{align}
This indeed captures the transition at $D = \kappa/4$ when $u \rightarrow 0$, see Fig. \ref{fig2}b. The phase transition is second order with critical exponent $\beta = 0.5$, suggesting that mean-field should be a good description \cite{LandauLifshitzStatPhys}. 

We denote the phase $D < \kappa/4$ as the $O(N)$ symmetric phase, and $D>\kappa/4$ as the symmetry-broken (SSB) phase, see Fig. \ref{fig2}a. Semiclassically, for $h = 0$, the symmetry-broken phase is characterized by a static vector on $S^{N-1}$. For $h \neq 0$, this vector undergoes persistent oscillation with frequencies set by $h$. In particular, for $N=2$ the SSB phase realizes a limit cycle, whereas for $N \geq 4$, it can realize not only limit cycles but also higher-dimensional limit tori (see Sec. \ref{sec_N_mode}).

%%%%%%%%%%%%%%%%%%%%%%%%%%%%%%%%%
%%%%%%%%%%%%%%%%%%%%%%%%%%%%%%%%%
\subsection{Steady-state Entanglement}

It is clear from the analytic and heuristic form of our dissipative steady state that there will be correlations between the $N$ modes that participate in the limit-cycle physics (i.e.,~their amplitudes must be correlated for them to lie on a sphere in phase space).  It is thus natural to ask whether these correlations are completely classical, or whether they correspond to a non-zero amount of entanglement between the modes.  Naively, one might expect the presence of steady-state entanglement for small-amplitude limit cycles, but that this entanglement would vanish in the semiclassical limit where the amplitude of the limit cycle becomes large.  

To study these questions in the simplest non-trivial setting, we specialize to the case of $N=2$.  Since the steady state is mixed, we quantify entanglement between the modes $\hat{a}_1$, $\hat{a}_2$ via the log negativity  $E_{N} = \log_2(||\rho^{T_{a_2}}_{ss}||_1)$, defined via the trace norm of the partially transposed density matrix \cite{PhysRevA.65.032314}.  We can calculate this directly from our analytic expression for the dissipative steady state.  Fixing $\kappa$, we study $E_N$ as a function of driving strength $D$ for several values of the nonlinearity $u$, see Fig. \ref{fig2}c. We find that $E_N$ increases with drive at small $D$, reaching a maximum near the limit cycle threshold, and then saturates to a finite value deep in the large photon number regime. To provide some context, we can compare the amount of steady state entanglement we have against a two-mode squeezed thermal state (TMST) that is constrained to have the same average total photon number and purity as our dissipative steady state.  The ratio of our entanglement to this reference TMST is also shown in Fig.~\ref{fig2}c; it exhibits a maximum value of $\sim 0.3$ as a function of $D / \kappa$, and decays logarithmically with increasing $D$.

While comparing against a TMST might seem natural, it is perhaps not a good choice of reference state.  While a TMST is necessarily Gaussian, the entanglement in our steady state comes purely from non-Gaussian correlations.  This follows immediately from the weak $O(N)$ symmetry, which restricts the covariance matrix to be proportional to $\delta_{ij}$ and thereby forces all steady-state intermode covariances, such as $\langle \hat{a}_1 \hat{a}_2\rangle$, to vanish. Hence, the non-zero entanglement we find is necessarily the consequence of non-zero, higher-order quantum correlations between the modes. 

It is also interesting to consider the steady-state entanglement across a different mode bipartition, e.g. between the $\hat{b}$ modes defined in Eq.~\eqref{b_mode_eq}.  In this mode basis, the system takes the form of a NDPO with Kerr nonlinearities, see App. \ref{app_b_entanglement}. Under this partition, $E_N$ no longer increases monotonically with drive. Instead, $E_N$ peaks near the threshold, then decreases and saturates to a slightly larger finite value.

\section{Phase diffusion}\label{sec_phase_diffusion}
We now turn from the steady-state properties to dynamics on the semiclassical attractor manifold. In the presence of quantum noise, the semiclassical motion acquires random diffusive motion along the attractor manifold. In the $O(N)$ case, we parameterize the attractor manifold by $N-1$ angular coordinates $\{\phi^\alpha\}$, and quantum noise leads to phase diffusion with diffusion matrix $D^{\alpha \beta}$ given by
\begin{align}
    \langle [\phi^\alpha(t) - \phi^\alpha(0)] [\phi^\beta(t) - \phi^\beta(0)] \rangle \sim 2 D^{\alpha \beta} t
\end{align}
Because drive is coherent, it does not introduce additional pump noise that appears in conventional incoherently pumped systems, and one might expect the resulting phase diffusion to be smaller. This indeed happens in some parameter regions of our model.

To make a connection with laser physics, we focus on the minimal $N=2$ case, where the model has $U(1)$ symmetry and only a single angular mode on the attractor manifold. In this case, the diffusion matrix becomes a diffusion coefficient and is directly related to the laser linewidth, which can be compared to the standard Schawlow-Townes limit. We show that with a careful choice of parameters, the diffusion coefficient can be reduced by a factor of two relative to the Schawlow-Townes limit, which we attribute to the absence of incoherent pump noise.

Our exact solution method for the quantum model unfortunately only yields steady-state quantities, and hence cannot directly yield dynamical quantities (e.g.~the spectrum of the Lindbladian, which could be used to infer the phase diffusion rate \cite{Ciuti_PRA_2018}).  As such, we consider the large photon number regime, where, similar to standard laser theory \cite{Zoller_textbook}, a semiclassical description can be used to describe how phase noise arises from quantum fluctuations.  In such an approach, one adds classical white noise to the semiclassical phase evolution equation Eq.~(\ref{phase_EOM}), with the strength of this noise chosen to mimic the effects of quantum noise.  We thus obtain a semiclassical Langevin equation for the phase of the form:
\begin{align}
    \partial_t \phi &= h + 4 u \sqrt{n_{ss}} \delta y_{\|} + \frac{\sqrt{\kappa}}{\sqrt{n_{\rm ss}}}\xi_{1},  \\
    \partial_t (\delta y_{\|}) &= -\kappa \delta y_{\|} + \sqrt{\kappa} \xi_2.
\end{align}
Here $\xi_j(t)$ are independent, real-valued zero-mean white noise processes, with autocorrelation functions
$\langle \xi_i(t) \xi_j(t') \rangle = \frac{1}{4}\delta_{ij}\delta(t-t')$ chosen to reproduce the correct quantum fluctuations. The normalization here corresponds to our convention for quadratures (i.e.,~vacuum noise yields a quadrature variance of $1/4$). 

Eq.~(\ref{phase_EOM}) shows that phase diffusion occurs both directly through the noise $\xi_1(t)$, but also through a coupling to the noise $y$-quadrature amplitude fluctuations (i.e.,~$\xi_2(t)$ also contributes).  As a result, the total phase diffusion contains two independent contributions, given by
\begin{align} \label{diffusion_coef}
    D_\phi = \frac{\kappa}{8n_{\rm ss}} + \frac{2 u^2 n_{\rm ss}}{\kappa}  = \frac{\kappa}{8n_{\rm ss}} \left(1 + \frac{16 u^2 n_{\rm ss}^2}{\kappa^2} \right).
\end{align}
Using the semiclassical total photon number $n_{\rm ss}$, we find $D_\phi = \frac{\kappa}{8n_{\rm ss}}r^2$ with $r = 4D/\kappa$. As the transition threshold is approached from above, $r \rightarrow 1^+$, the phase diffusion approaches 
\begin{equation}
    D_{\phi} \rightarrow \frac{\kappa}{8n_{\rm ss}}
\end{equation}
%$D_{\phi} \rightarrow \kappa/8 n_{\rm ss}$. 
Expressed in units of $\kappa/n_{\rm ss}$, this is a factor of two below the standard quantum limit on the the phase diffusion (or linewidth) of a laser oscillator, i.e.,~the Schawlow-Townes limit \cite{Courtois_1991_OPO}. At a heuristic level, this factor of two reduction can be understood intuitively.  For a standard incoherently-pumped laser or limit cycle oscillator, vacuum fluctuations from both the loss reservoir and the pumping reservoir contribute equally to the phase diffusion.  In our setup we eliminate one of these sources of noise, as the pumping is now fully coherent, hence a factor of two improvement.  The situation is more nuanced than this however, because of the amplitude-phase coupling inherent in Eq.~(\ref{phase_EOM}).  

The above regime of reduced phase diffusion is achieved just above threshold, requiring a drive amplitude $D \rightarrow \kappa/4$.  Note that one can still independently tune the size of the limit cycle and the total average photon number, as this depends on the size of the nonlinearity $u$:  $n_{\rm ss} \sim D/u$. To access the regime of reduced phase diffusion, one should operate in a regime near the transition threshold with weak nonlinearity, which is achieved using a hierarchy of scales $u \ll D \lesssim \kappa$. Coupling the phase mode to an output field then realizes a laser with linewidth $\Gamma = 2D_\phi = \kappa/4n_{\rm ss}$, which is half the Schawlow–Townes linewidth of a conventional laser \cite{Wiseman_laser_SQL}.

From basic number-phase uncertainty, one might expect that the reduced phase noise of our limit cycle will result in enhanced amplitude or number fluctuations, i.e., fluctuations of the $x$ quadratures that are normal to the limit cycle.  The impact of these fluctuations is conventionally characterized by the steady-state Fano ratio, which is the normalized variance of the photon number.  Using the exact solution, we find 
\begin{align}
    \text{Fano} = \frac{\text{Var}(\hat{n})}{\text{Mean}(\hat{n})} = \frac{\text{Var}(\hat{m})}{\langle \hat{m} \rangle} + \frac{1}{2},
\end{align}
where $m$ is the number of pair excitations in the doubled system with $\text{Var}(\hat{m}) = \langle \hat{m}(\hat{m}-1) \rangle + \langle \hat{m} \rangle - \langle \hat{m} \rangle^2$.  The average and variance of these pair excitations can be computed exactly:
\begin{align}
    \langle \hat{m} \rangle &= \frac{\lambda}{\delta \delta^*}\frac{{}_1 F_2(2; \delta+1, \delta^*+1; \lambda)}{{}_1 F_2(1; \delta, \delta^*; \lambda)} \\
    \langle \hat{m}(\hat{m}-1)\rangle &= \frac{2\lambda^2}{\delta \delta^* (\delta+1)(\delta^*+1)} \frac{{}_1 F_{2}(3; \delta+2, \delta^*+2; \lambda)}{{}_1 F_{2}(1; \delta, \delta^*; \lambda)} 
\end{align}
Recall that for a simple coherent state, the Fano ratio would be $1$.  We find using the above expressions that in the limit of $u \rightarrow 0$, the above expression converges to $\text{Fano} \rightarrow 1 + \frac{1}{2}\frac{1}{(4D/\kappa)^2-1}$ above threshold, see App. \ref{app_obs}. Thus, as $D \rightarrow \kappa/4$, the Fano ratio diverges, indicating strongly enhanced amplitude fluctuation near the limit-cycle threshold, see Fig. \ref{fig2}d. 

Remarkably, while the semiclassical prediction of the mean photon number agrees with the exact solution only when $\langle \hat{n} \rangle \gtrsim \mathcal{O}(10^3)$, the Fano ratio calculated from semiclassics agrees well with the exact solution even when $\langle \hat{n} \rangle \sim \mathcal{O}(1)$. For strong driving $D > \kappa$, the photon statistics quickly approach Poissonian limit ($F \rightarrow$ 1), consistent with the emergence of large photon coherent state. The phase diffusion in Eq. \ref{diffusion_coef} is minimized when drive strength $D = \kappa/(2\sqrt{2})$, where $D_{\phi, \rm min} = \frac{\kappa}{4n_{\rm ss}}$, and at this point the Fano ratio is $F = 1.5$. The same value  also appears in the weak drive limit $D \rightarrow 0$. For comparison, the steady state of the quantum van der Pol oscillator in the large photon number limit also has Fano ratio $F = 1.5$. In our model, the large photon number limit can be reached either by sending $u \rightarrow 0$ or $D \rightarrow \infty$, and these two limits lead to different steady-state behavior. In the weak nonlinearity limit $u \rightarrow 0$, the Fano ratio approaches $F \rightarrow 1 + \frac{1}{2[(4D/\kappa)^2-1]}$ above threshold, and gives $F=1.5$ at $D = \kappa/(2\sqrt{2})$. By contrast, in the strong drive limit, the Fano ratio approaches $F \rightarrow 1$, indicating smaller amplitude fluctuations compared to the quantum van der Pol model, see App. \ref{app_van}.

Our focus here has been on unavoidable fluctuations in the limit cycle state (both phase and amplitude) arising from the vacuum fluctuations in dissipative reservoirs.  We note that a very different phase-noise suppression mechanism was studied in Ref.~\cite{Roukes_PRL_2012}, in a fully classical version of our $N=2$ setup.  The focus there was on classical noise arising from fluctuations in the parametric drive frequency.  The noise-reducing mechanism they find is unrelated to the factor-of-two suppression of phase noise due to quantum noise described above.
We also note that phase noise due to quantum fluctuations has been studied in a two-mode NDPO setup that unlike our system has a dissipative nonlinearity stemming from pump-depletion effects \cite{Courtois_1991_OPO}; in contrast, we have a fully Hamiltonian, Kerr-type nonlinearity.  While the setup in Ref.~\cite{Courtois_1991_OPO} does not have amplitude-phase coupling, it also exhibits a suppression of phase noise by a factor of two below the usual Schawlow-Townes limit.

\vspace{0.5 cm}

\section{Dynamics of the $N$-mode model}\label{sec_N_mode}

\subsection{Basic features of the $O(N)$ limit cycle phase}

We now turn to the new kinds of dynamical phenomena that arises for $N>2$ in the $O(N)$ model introduced in 
Eq.~\eqref{n_mode_model}. As in the two-mode case, the semiclassical attractor manifold is determined by a constraint on the amplitude of the $x$-quadratures. For the $N$ modes, one obtain
\begin{align}
    \sum_{i=1}^N x_i^2 = n_{\rm ss},
\end{align}
which defines an $(N-1)$-sphere $S^{N-1}$. 

Although the antisymmetric matrix $K$ does not change the quantum steady-state, it modifies the late-time dynamics as it approaches the steady-state, which will be our focus here. In particular, the antisymmetric coupling $hK$ generates persistent oscillation on $S^{N-1}$, giving rise to oscillatory states that are delocalized over the $N$ modes. The $y$-quadrature amplitudes all vanish on the attractor manifold, $y_i = 0$. As in the two-mode case, linearizing around the semiclassical steady state yields $N-1$ undamped (gapless) modes, corresponding to tangential motion along the steady state manifold. In contrast, the single radial fluctuation away from the limit cycle involving only $x$-quadrature is massive (i.e.,~decays exponentially), as are the $y$-quadrature fluctuations. The $N-1$ gapless modes are the multimode generalization of the single phase mode of the $U(1)$ limit cycle. We note that a different kind of non-equilibrium $O(N)$ limit cycle model was recently studied in Ref.~\cite{Diehl_PRX_2024} \footnote{
In \cite{Diehl_PRX_2024},the plane of rotation is spontaneously chosen in the limit cycle phase, which corresponds to a different symmetry-breaking pattern than our model. In our case, the semiclassical attractor is a sphere $S^{N-1}$, and the antisymmetric coupling matrix $K$ generates motion along the manifold. The plane of oscillation is therefore fixed explicitly by $K$, rather than being selected spontaneously. }.

 Parameterizing the sphere $S^{N-1}$ by angular coordinates $\phi^\alpha$, with $\alpha = 1, \cdots, N-1$, we can decompose fluctuations in the local tangent basis on the sphere. Recall that in the case of $N=2$ (see Eq.~\eqref{phase_EOM}), there is only a single $\phi$ that corresponds to the tangential direction on $S^1$, whose motion couples to the fluctuation of $y$-quadrature tangential to the limit cycle. For general $N$, each $\phi^\alpha$ labels a direction on $S^{N-1}$, and $\delta y^\alpha$ denotes the component of the $y$-quadrature fluctuations along that direction. The linearized dynamics then take the form (see App. \ref{app_stability}):
 \begin{align}
     \partial_t \phi^\alpha = \omega^\alpha + 4u \sqrt{n_{\rm ss}} \delta y^\alpha \qquad 
     \partial_t (\delta y^\alpha) = - \kappa \delta y^\alpha
 \end{align}
 The tangential $y$-quadrature fluctuations are damped, and the $N-1$ angular variables govern the long-time dynamics on the attractor manifold.

In practice, the explicit form of $\omega^\alpha$ is complicated, and the dynamics are more transparent after bringing $K$ to a canonical block form. For a general antisymmetric matrix $K$, one can perform an orthogonal transformation $O \in O(N)$ such that 
$K = O^T \Sigma O$, where 
\begin{align}
    \Sigma = \bigoplus_{r=1}^{\lfloor N/2 \rfloor} \lambda_r (i\sigma_y) \oplus \begin{cases}
        0 \qquad \text{if N is odd}\\
        \emptyset \qquad \text{if N is even}
    \end{cases}
\end{align}
We define new modes $\vb{b} = O\vb{a}$, and write $\vb{b} = e^{i\theta}(\vb{q}+i\vb{p}) $, where $\vb{q}, \vb{p}$ are real $N$-component vectors. In this basis, the antisymmetric matrix $K$ is block diagonal, and the dynamics decomposes into  $\lfloor N/2 \rfloor$ independent two-dimensional planes, each associated with one $2 \times 2$ blocks of $K$. Within each of the $r$-th plane spanned by $\vb{q}^{(r)} = (q_{2r-1}, q_{2r})^T, \vb{p}^{(r)} = (p_{2r-1}, p_{2r})^T$, where $r = 1, \cdots, \lfloor N/2 \rfloor$, the semiclassical equation of motion are
\begin{align}
    \partial_t \vb{q}^{(r)} &= (2un + 2D \cos(2\theta)) \vb{p}^{(r)} + h \lambda_r J \vb{q}^{(r)} \\
    \partial_t \vb{p}^{(r)} &= (-2un + 2D \cos(2\theta)) \vb{q}^{(r)} - \kappa \vb{p}^{(r)} + h\lambda_r J \vb{p}^{(r)}
\end{align}
where $n$ is the total photon number, $\theta = \frac{1}{2}\arcsin(\kappa/4D)$, and $J  = -i\sigma_y$ generates rotation within the 2D plane. These equations are the generalization of the two-mode quadrature equations in 
Eq.~\eqref{quadrature_EOM}. In the semiclassical steady-state, the vector $\vb{q}^{(r)}$ in the $r$-th plane precesses at frequency $h\lambda_r$ subject to a fixed length $q_{2r-1}^2 + q_{2r}^2 = \rho_r^2$. Therefore, the oscillatory dynamics in the $O(N)$ model reduce to a collection of coupled amplitude modes together with several independent phase rotation in each of the 2 $\times$ 2 subspace. 

\begin{figure}
    \centering
    \includegraphics[width=\linewidth]{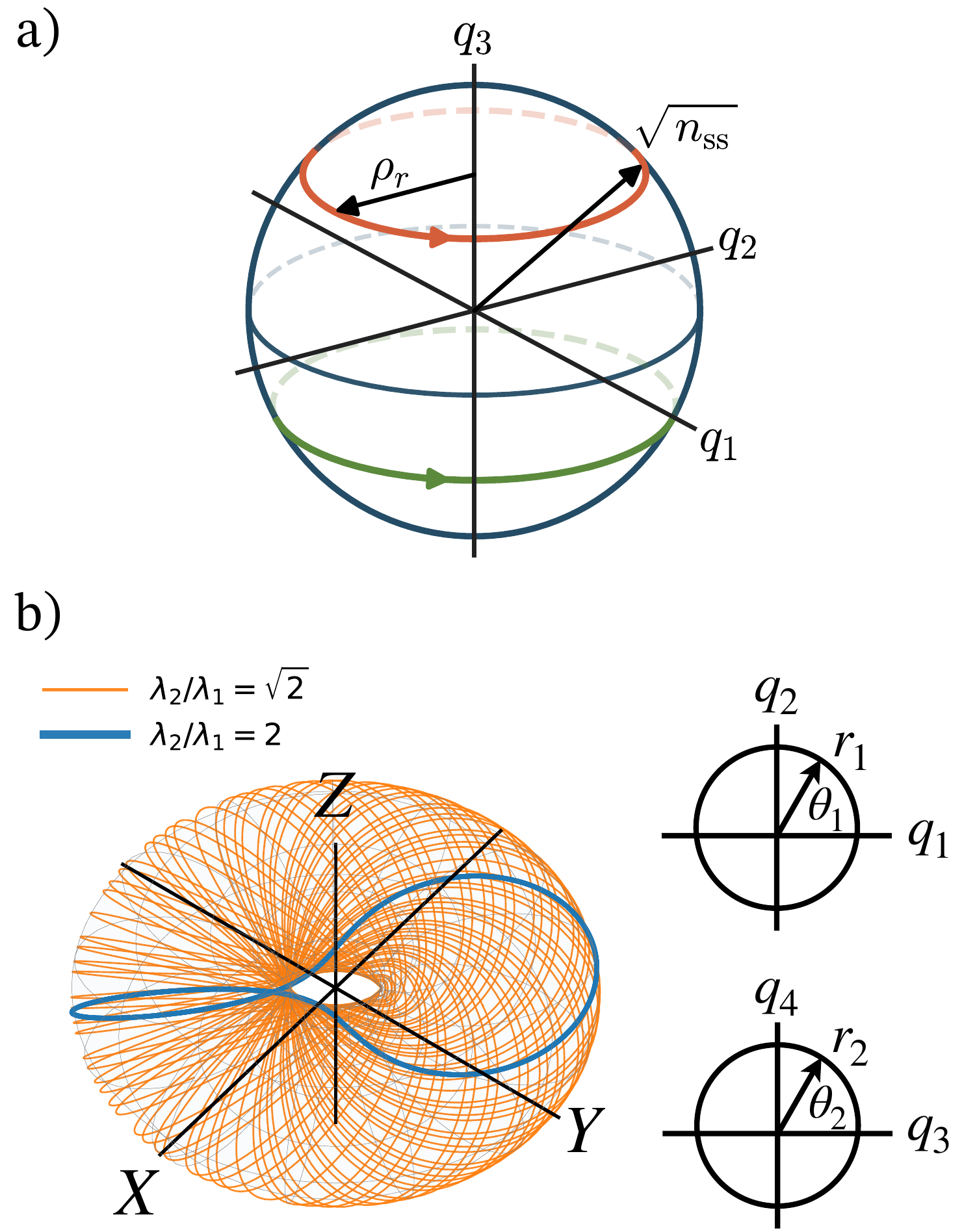}
    \caption{a) Attractor manifold for $N=3$ with antisymmetric coupling in the first two modes, in coordinates that block diagonalizes $K$. Red and green are two realizations on the semiclassical steady-state solution, depending on the initial condition. b) Stereographic projection of a Clifford torus $T^2 \subset S^3$ into $\mathbb{R}^3$, parameterized by $(r_1 e^{i\theta_1}, r_2 e^{i\theta_2})$ and subject to $r_1^2 + r_2^2 = n_{\rm ss}$. Left: when $\lambda_2/\lambda_1$ is irrational, the trajectory is quasiperiodic and fills the torus; when it is rational, the trajectory closes onto a periodic orbit. Right: the corresponding circular motion in 2 $\times$ 2 blocks with radii $r_i$ and angular velocity $h \lambda_i$.}
    \label{fig3}
\end{figure}

As in the two-mode case, we can parameterize the vector $\vb{q}^{(r)}$ in polar form
\begin{align}
    \vb{q}^{(r)} = \rho_r 
    \begin{pmatrix}
    \cos(\theta_r) \\
    \sin(\theta_r)
    \end{pmatrix},
\end{align}
where $\rho_r$ is the radius of $S^{N-1}$ projected to the $r$-th plane, and $\theta_r$ is the angular variable. These $\rho_r$ are not independent, and they must satisfy the overall amplitude constraint
\begin{align}\label{norm_constraint}
    \sum_{r=1}^{\lfloor N/2 \rfloor} \rho_r^2 + \varsigma^2 = n_{\rm ss}.
\end{align}
Here $\varsigma$ is the leftover coordinate when $N$ is odd, and it is zero when $N$ is even. This constraint expresses the fact that the total radius on the $S^{N-1}$ manifold is fixed, while the amplitude $\rho_r$ determines how the radius is distributed among the different rotation planes. Fig.~\ref{fig3}a shows the resulting attractor manifold for $N=3$ mode.

For each of the $r$-th block, the steady-state solution is $\vb{q}^{(r)}_0 = \rho_r \hat{\vb{r}}$. We decompose fluctuations around it as
\begin{align}
    \delta \vb{q}^{(r)} &= \delta \rho_{\perp,r} \hat{\vb{r}} + \rho_r \theta_r \hat{\vb{s}} \\
    \delta \vb{p}^{(r)} &= p_{\perp,r} \hat{\vb{r}} + p_{\|, r}\hat{\vb{s}}
\end{align}
where $\hat{\vb{r}}, \hat{\vb{s}}$ are the radial and tangential unit vectors to the circle on the $r$-th plane, respectively. This is the analogue of the decomposition of fluctuation into radial and tangential directions on a circle in the two-mode case. The linearized equation of motion yields
\begin{align}
    \partial_t \theta_r = h\lambda_r + \frac{4 u n_{\rm ss}}{\rho_r} p_{\|, r}, \qquad \partial_t \rho_r = 4 u n_{\rm ss} p_{\perp, r}
\end{align}
while both $p_{\|, r}$ and $p_{\perp, r}$ decay at rate $\kappa$ (see App. \ref{app_stability}). Therefore, in the long-time limit, the radii $\rho_r$ settle to a fixed value subject to the constraint in Eq.~\eqref{norm_constraint}, which selects a particular circle within the corresponding 2D plane, while the phase $\theta_r$ exhibits uniform oscillation along the circle with frequency $h \lambda_r$.

In this way, the $N-1$ gapless modes naturally split into two kinds: $\lfloor N/2 \rfloor$ phase modes $\theta_r$ describing rotations in each of the 2 $\times$ 2 blocks, and  $N-1-\lfloor N/2 \rfloor$ amplitude modes describing how the fixed total radius is shared among different blocks. In the presence of noise, the amplitude mode redistributes the fixed total radius among different $2 \times 2$  blocks while staying on the $S^{N-1}$ manifold, and the phase fluctuation changes the angle inside each block. 

\subsection{Emergence of quantum limit tori}

For $N\geq 4$, our model can support several independent oscillation frequencies and therefore realizes a \textit{quantum limit tori} rather than a simple limit cycle \cite{Nowoczyn2025Universal}. When these frequencies are commensurate, the trajectory eventually closes and the motion is periodic with a well-defined period. By contrast, when they are incommensurate, the trajectory never closes and the system exhibits quasiperiodic motion on an invariant torus $T^n = (S^1)^n$, where $n = \lfloor N/2 \rfloor$, see Fig. \ref{fig3}b. Each block contributes one independent angular variable oscillating with frequency $h\lambda_r$, and the torus is simply the product of these independent circular motions. The resulting torus is embedded in $S^{N-1}$, which realizes the geometry of a generalized Clifford torus.

Starting from an initial condition, the massive fluctuations decay first, and the system relaxes towards the attractor manifold. Initial condition determines the how the total radius $\sqrt{n_{\rm ss}}$ is distributed among the different $\rho_r$, which determines a particular torus $T^{n}$ embedded in $S^{N-1}$. The frequency of oscillation is directly set by the structure of the coupling matrix $K$. In this sense, the geometry of the motion is highly tunable: by simply changing the coupling matrix, one can change the set of frequencies $h \lambda_r$, and hence control whether the resulting motion is periodic or quasiperiodic.

In the presence of quantum noise, fluctuations induce slow diffusion along the $N-1$ gapless direction on the attractor manifold. In the $U(1)$ case, noise drives the single angular mode to diffuse along the circle, and the dynamics is characterized by a single diffusion coefficient. For a higher-dimensional manifold, quantum noise not only changes the angular variable $\theta_r$, but also slowly redistributes the amplitude $\rho_r$ subject to the total radius constraint. Therefore, the resulting dynamics on the manifold is captured by a diffusion matrix. Quantum fluctuations cause the system to wander not only along a given semiclassical torus, but also between nearby tori compatible with the total radius constraint. At finite $n_{\rm ss}$, this diffusion eventually restores the full $O(N)$ symmetry at long times, with rate vanishing as $1/n_{\rm ss}$ in the thermodynamic limit.

\section{Symmetry breaking terms and connections to other limit cycle models based on parametric drives}

Our general $O(N)$ symmetric model provides a unifying framework for understanding how parametric driving in a much wider class of models can give rise to limit cycles.  As we show here, the same basic mechanism we describe still holds even if the precise symmetry in our model is weakly broken.  The symmetry breaking terms lead to the limit cycle phase only existing for a finite range of drive amplitudes. For large amplitudes, these terms give rise to a phase locking effect.  We discuss this more below for the simplest $N=2$ case, and show that previously studied parametrically driven models that gave limit cycles without $U(1)$ symmetry \cite{bello_persistent_2019, calvanese_strinati_theory_2019} can be understood as symmetry-broken versions of our model.  

% is that it provides a unifying framework for understanding how parametric drives can give rise to continuous symmetry and limit cycle behaviors. In particular, the $N=2$ mode model provides a simple way of understanding previously studied parametrically driven models without $U(1)$ symmetry \cite{bello_persistent_2019}: they can be understood as a symmetry-broken version of the model studied here. 

To be concrete, we perturb the nonlinear interaction in our $N=2$ model with a term that explicitly breaks the $U(1)$ symmetry:
\begin{align}
    \hat{H}_{\rm NL} = u(\hat{n}_1 + \hat{n}_2)^2 - 2 \, \delta u \, \hat{n}_1 \hat{n}_2
\end{align}
Here, $\delta u$ controls the strength of the symmetry-breaking perturbation. The $\delta u$ term corresponds to a mismatch between self-Kerr and cross-Kerr interactions.  
Its effect is to introduce a phase-locking term in the equation of motion of the phase variable. For weak $\delta u$, the leading-order phase dynamics (i.e., Eq. \eqref{phase_EOM}) is modified and becomes an Adler-type equation (see App. \ref{app_symm_break}):
\begin{align}
     \partial_t \phi = h  + \frac{2 u \, \delta u \,}{\kappa} n_{\rm ss}^2 \sin(4\phi)
\end{align}

We now have a simple way to understand the effect of $\delta u$.  When the product of this perturbation and the bare limit cycle size $n_{\rm ss}^2$ is small compared to the oscillation frequency, the phase variable can still wind continuously, and the system retains limit cycle dynamics. In contrast, for a sufficiently large drive, the phase-locking term dominates and pins the phase, thus destroying the limit cycle. Since the total photon number scales as $n_{\rm ss} \propto D/u$, the boundary between phase locking and limit cycles is given by
\begin{align}
    h \propto \frac{\delta u}{u \kappa} D^2
\end{align}

This discussion shows that our symmetric model can be understood as a kind of parent model in which parametric drive and symmetry generators give rise to limit cycle behavior. In the symmetric case, an arbitrarily strong drive can lead to robust limit cycle behavior with any finite $h$. With symmetry-broken interactions, driving generates a phase-locking term, producing a competition between phase winding and phase locking. The persistence of limit cycle behavior over a finite range of oscillation frequencies, even in the presence of symmetry-breaking interactions, demonstrates the robustness of our model against experimental imperfections.

\section{Discussion}
Our work presents a symmetry-based principle for constructing a broad class of exactly-solvable, parametrically-driven multimode bosonic systems exhibiting robust limit cycles behavior. The key observation is that while parametric drives break the $U(1)$ phase symmetry of each individual mode, the full dynamics can still retain a weak $O(N)$ symmetry in mode space. This allows parametric drive and continuous symmetry to coexist in a surprisingly simple and experimentally relevant setting.

Semiclassically, the dynamics quickly relaxes towards an $S^{N-1}$ attractor manifold, which spontaneously breaks the $O(N)$ symmetry. Motion normal to the manifold is damped, while the $N-1$ tangential directions are gapless and govern the long-time dynamics. At the full quantum level, the Lindbladian admits a unique steady-state density matrix, which restores the $O(N)$ symmetry. Therefore, the dynamics exhibits a separation of scales: trajectories quickly settle towards the attractor manifold and then undergo slow phase diffusion along it due to quantum noise. In the two-mode case, this process is described by a single phase diffusion, with diffusion coefficient scaling as $D_\phi \sim 1/n_{\rm ss}$. For general $N$, this late-time process is characterized by a diffusion matrix on the attractor manifold.

Another central result of our work is that the antisymmetric coupling matrix $K$ reshapes the late-time dynamics without changing the steady-state density matrix. The matrix $K$ produces a persistent oscillation along the attractor manifold, giving rise to a variety of dynamical phenomena, such as limit cycle and limit tori, depending on the choice of $K$. This richness in dynamics is not captured by the steady-state density matrix alone but also in how the system approaches the steady-state at late-time. It would therefore be very interesting in future works to develop techniques that study this quantum dynamics beyond the semiclassical description.

Another interesting direction for future work would be to study the properties of the output field.  The $O(N)$ symmetric model studied here provides a natural starting point to engineer
an unusual kind of 
coherent multimode light source.  The coherence of the outgoing light would be characterized (like our limit cycle) by an order parameter living in $S^{N-1}$, and this coherence would be effectively delocalized over many modes.  Since the internal modes supporting the limit cycle support gapless collective modes associated with broken $O(N)$ symmetry at the semiclassical level, one may expect the emitted light exhibit strong multimode entanglement and high-order quantum correlations inherited from the underlying symmetry structure. 

 Our model is directly realizable in a variety of superconducting circuit platforms and provides a minimal setting in which robust limit cycle with continuous symmetry can be realized via purely coherent drive. The dynamics of each bosonic mode can be monitored by weakly coupling it to a waveguide and measuring the outfield field using homodyne or heterodyne detection \cite{Clerk2010}. It thus provides a natural starting point to study a variety of other interesting phenomena, such as quantum synchronization, and phase entrainment in the presence of small symmetry-breaking perturbations.

\section{Acknowledgments}
We thank Andrew Pocklington, Sergei Shmakov, Cheyne Weis and Jamir Marino for helpful discussions.
This work was supported by the 
Army Research Office under grant W911NF-25-1-0286 
and the Simons Foundation through a Simons Investigator award (Grant No. 669487).

\bibliography{bibliography.bib}

@article{two_photon_exact,
  title = {Competition between Two-Photon Driving, Dissipation, and Interactions in Bosonic Lattice Models: An Exact Solution},
  author = {Roberts, David and Clerk, A. A.},
  journal = {Phys. Rev. Lett.},
  volume = {130},
  issue = {6},
  pages = {063601},
  numpages = {6},
  year = {2023},
  month = {Feb},
  publisher = {American Physical Society},
  doi = {10.1103/PhysRevLett.130.063601},
  url = {https://link.aps.org/doi/10.1103/PhysRevLett.130.063601}
}

@article{Wiseman_laser_SQL,
  title = {Light amplification without stimulated emission: Beyond the standard quantum limit to the laser linewidth},
  author = {Wiseman, H. M.},
  journal = {Phys. Rev. A},
  volume = {60},
  issue = {5},
  pages = {4083--4093},
  numpages = {0},
  year = {1999},
  month = {Nov},
  publisher = {American Physical Society},
  doi = {10.1103/PhysRevA.60.4083},
  url = {https://link.aps.org/doi/10.1103/PhysRevA.60.4083}
}

@article{hTRS,
  title = {Hidden Time-Reversal Symmetry, Quantum Detailed Balance and Exact Solutions of Driven-Dissipative Quantum Systems},
  author = {Roberts, David and Lingenfelter, Andrew and Clerk, A.A.},
  journal = {PRX Quantum},
  volume = {2},
  issue = {2},
  pages = {020336},
  numpages = {33},
  year = {2021},
  month = {Jun},
  publisher = {American Physical Society},
  doi = {10.1103/PRXQuantum.2.020336},
  url = {https://link.aps.org/doi/10.1103/PRXQuantum.2.020336}
}

@article{bello_persistent_2019,
  title = {Persistent Coherent Beating in Coupled Parametric Oscillators},
  author = {Bello, Leon and Calvanese Strinati, Marcello and Dalla Torre, Emanuele G. and Pe'er, Avi},
  journal = {Phys. Rev. Lett.},
  volume = {123},
  issue = {8},
  pages = {083901},
  numpages = {6},
  year = {2019},
  month = {Aug},
  publisher = {American Physical Society},
  doi = {10.1103/PhysRevLett.123.083901}
}

@article{Vincenzo_Nature_2025,
	title = {Adaptive locomotion of active solids},
	volume = {639},
	issn = {1476-4687},
	number = {8056},
	urldate = {2026-01-14},
	journal = {Nature},
	publisher = {Nature Publishing Group},
	author = {Veenstra, Jonas and Scheibner, Colin and Brandenbourger, Martin and Binysh, Jack and Souslov, Anton and Vitelli, Vincenzo and Coulais, Corentin},
	month = mar,
	year = {2025},
	pages = {935--941},
    doi = {10.1038/s41586-025-08646-3}
}

@article{calvanese_strinati_theory_2019,
  title = {Theory of coupled parametric oscillators beyond coupled Ising spins},
  author = {Calvanese Strinati, Marcello and Bello, Leon and Pe'er, Avi and Dalla Torre, Emanuele G.},
  journal = {Phys. Rev. A},
  volume = {100},
  issue = {2},
  pages = {023835},
  numpages = {17},
  year = {2019},
  month = {Aug},
  publisher = {American Physical Society},
  doi = {10.1103/PhysRevA.100.023835}
}

@article{Bruder_PRL_2018,
  title = {Synchronizing the Smallest Possible System},
  author = {Roulet, Alexandre and Bruder, Christoph},
  journal = {Phys. Rev. Lett.},
  volume = {121},
  issue = {5},
  pages = {053601},
  numpages = {5},
  year = {2018},
  month = {Jul},
  publisher = {American Physical Society},
  doi = {10.1103/PhysRevLett.121.053601},
}

@Article{Buca_2022_SciPost,
	title={{Algebraic theory of quantum synchronization and limit cycles under dissipation}},
	author={Berislav Buča and Cameron Booker and Dieter Jaksch},
	journal={SciPost Phys.},
	volume={12},
	pages={097},
	year={2022},
	publisher={SciPost},
	doi={10.21468/SciPostPhys.12.3.097},
	url={https://scipost.org/10.21468/SciPostPhys.12.3.097},
}

@article{Diehl_PRX_2024,
  title = {Universal Phenomenology at Critical Exceptional Points of Nonequilibrium $\mathrm{O}(N)$ Models},
  author = {Zelle, Carl Philipp and Daviet, Romain and Rosch, Achim and Diehl, Sebastian},
  journal = {Phys. Rev. X},
  volume = {14},
  issue = {2},
  pages = {021052},
  numpages = {44},
  year = {2024},
  month = {Jun},
  publisher = {American Physical Society},
  doi = {10.1103/PhysRevX.14.021052}
}

@article{schawlow_infrared_1958,
	title = {Infrared and {Optical} {Masers}},
	volume = {112},
	copyright = {http://link.aps.org/licenses/aps-default-license},
	issn = {0031-899X},
	url = {https://link.aps.org/doi/10.1103/PhysRev.112.1940},
	doi = {10.1103/PhysRev.112.1940},
	 
	number = {6},
	urldate = {2026-01-14},
	journal = {Physical Review},
	author = {Schawlow, A. L. and Townes, C. H.},
	month = dec,
	year = {1958},
	pages = {1940--1949}
}

@article{PhysRevA.108.062216,
  title = {Entangled time-crystal phase in an open quantum light-matter system},
  author = {Mattes, Robert and Lesanovsky, Igor and Carollo, Federico},
  journal = {Phys. Rev. A},
  volume = {108},
  issue = {6},
  pages = {062216},
  numpages = {13},
  year = {2023},
  month = {Dec},
  publisher = {American Physical Society},
  doi = {10.1103/PhysRevA.108.062216}
}

@article{PhysRevA.65.032314,
  title = {Computable measure of entanglement},
  author = {Vidal, G. and Werner, R. F.},
  journal = {Phys. Rev. A},
  volume = {65},
  issue = {3},
  pages = {032314},
  numpages = {11},
  year = {2002},
  month = {Feb},
  publisher = {American Physical Society},
  doi = {10.1103/PhysRevA.65.032314},
  url = {https://link.aps.org/doi/10.1103/PhysRevA.65.032314}
}

@article{Haque_PRL_2025,
  title = {Quantum Origin of Limit Cycles, Fixed Points, and Critical Slowing Down},
  author = {Dutta, Shovan and Zhang, Shu and Haque, Masudul},
  journal = {Phys. Rev. Lett.},
  volume = {134},
  issue = {5},
  pages = {050407},
  numpages = {7},
  year = {2025},
  month = {Feb},
  publisher = {American Physical Society},
  doi = {10.1103/PhysRevLett.134.050407},
  url = {https://link.aps.org/doi/10.1103/PhysRevLett.134.050407}}

@article{Cross_PRR_2021,
	title = {Quantum limit cycles and the {Rayleigh} and van der {Pol} oscillators},
	volume = {3},
	issn = {2643-1564},
	number = {1},
	urldate = {2026-01-14},
	journal = {Physical Review Research},
	author = {Ben Arosh, Lior and Cross, M. C. and Lifshitz, Ron},
	month = feb,
	year = {2021},
   doi = {10.1103/PhysRevResearch.3.013130}
}

@article{Cosme_PRA_2024,
  title = {Realizing limit cycles in dissipative bosonic systems},
  author = {Skulte, Jim and Kongkhambut, Phatthamon and Ke\ss{}ler, Hans and Hemmerich, Andreas and Mathey, Ludwig and Cosme, Jayson G.},
  journal = {Phys. Rev. A},
  volume = {109},
  issue = {6},
  pages = {063317},
  numpages = {10},
  year = {2024},
  month = {Jun},
  publisher = {American Physical Society},
  doi = {10.1103/PhysRevA.109.063317}
}

@article{Lindblad1976,
  author    = {Lindblad, G{\"o}ran},
  title     = {On the Generators of Quantum Dynamical Semigroups},
  journal   = {Communications in Mathematical Physics},
  volume    = {48},
  number    = {2},
  pages     = {119--130},
  year      = {1976},
  doi       = {10.1007/BF01608499}
}

@article{Gorini:1975nb,
    author = "Gorini, Vittorio and Kossakowski, Andrzej and Sudarshan, E. C. G.",
    title = "{Completely Positive Dynamical Semigroups of N Level Systems}",
    reportNumber = "CPT-244-TEXAS, ORO-3992-200",
    doi = "10.1063/1.522979",
    journal = "J. Math. Phys.",
    volume = "17",
    pages = "821",
    year = "1976"
}

@article{Nowoczyn2025Universal,
  title={Universal quantum melting of quasiperiodic attractors in driven-dissipative cavities},
  author={Nowoczyn, Caroline and Mathey, Ludwig and Seibold, Kilian},
  journal={arXiv preprint arXiv:2507.03764},
  year={2025},
  archivePrefix={arXiv},
  primaryClass={quant-ph}
}

@article{Scheibner2020,
  title   = {Odd elasticity},
  author  = {Scheibner, Colin and Souslov, Anton and Banerjee, Debarghya and Sur{\'o}wka, Piotr and Irvine, William T. M. and Vitelli, Vincenzo},
  journal = {Nature Physics},
  year    = {2020},
  volume  = {16},
  number  = {4},
  pages   = {475-480},
  doi     = {10.1038/s41567-020-0795-y}
}

@article{CQA,
  author  = {K. Stannigel and P. Rabl and P. Zoller},
  title   = {Driven-dissipative preparation of entangled states in cascaded quantum-optical networks},
  journal = {New Journal of Physics},
  volume  = {14},
  number  = {6},
  pages   = {063014},
  year    = {2012},
  doi     = {10.1088/1367-2630/14/6/063014}
}

@article{Zalatel_RMP_2023,
  title = {Colloquium: Quantum and classical discrete time crystals},
  author = {Zaletel, Michael P. and Lukin, Mikhail and Monroe, Christopher and Nayak, Chetan and Wilczek, Frank and Yao, Norman Y.},
  journal = {Rev. Mod. Phys.},
  volume = {95},
  issue = {3},
  pages = {031001},
  numpages = {34},
  year = {2023},
  month = {Jul},
  publisher = {American Physical Society},
  doi = {10.1103/RevModPhys.95.031001},
  url = {https://link.aps.org/doi/10.1103/RevModPhys.95.031001}
}

@article{Clerk_PRX_2014,
  title = {Laser Theory for Optomechanics: Limit Cycles in the Quantum Regime},
  author = {L\"orch, Niels and Qian, Jiang and Clerk, Aashish and Marquardt, Florian and Hammerer, Klemens},
  journal = {Phys. Rev. X},
  volume = {4},
  issue = {1},
  pages = {011015},
  numpages = {21},
  year = {2014},
  month = {Jan},
  publisher = {American Physical Society},
  doi = {10.1103/PhysRevX.4.011015},
  url = {https://link.aps.org/doi/10.1103/PhysRevX.4.011015}
}

@book{Zoller_textbook,
  author    = {Crispin Gardiner and Peter Zoller},
  title     = {Quantum Noise: A Handbook of Markovian and Non-Markovian Quantum Stochastic Methods with Applications to Quantum Optics},
  publisher = {Springer},
  address   = {Berlin, Heidelberg},
  year      = {2004},
  edition   = {3},
  series    = {Springer Series in Synergetics},
  isbn      = {978-3-540-22301-6}
}

@book{scully_zubairy_1997,
  author    = {Marlan O. Scully and M. Suhail Zubairy},
  title     = {Quantum Optics},
  year      = {1997},
  publisher = {Cambridge University Press},
  address   = {Cambridge},
  edition   = {1}
}

@article{else_2020_review,
  author  = {Dominic V. Else and Christopher Monroe and Chetan Nayak and Norman Y. Yao},
  title   = {Discrete Time Crystals},
  journal = {Annual Review of Condensed Matter Physics},
  volume  = {11},
  pages   = {467--499},
  year    = {2020},
  doi     = {10.1146/annurev-conmatphys-031119-050658}
}

@article{Sacha_2018_review,
  author  = {Krzysztof Sacha and Jakub Zakrzewski},
  title   = {Time crystals: a review},
  journal = {Reports on Progress in Physics},
  volume  = {81},
  number  = {1},
  pages   = {016401},
  year    = {2018},
  doi     = {10.1088/1361-6633/aa8b38}
}

@article{Thompson_Nature_2012,
  author  = {Justin G. Bohnet and Zilong Chen and Joshua M. Weiner and Dominic Meiser and Murray J. Holland and James K. Thompson},
  title   = {A steady-state superradiant laser with less than one intracavity photon},
  journal = {Nature},
  volume  = {484},
  pages   = {78--81},
  year    = {2012},
  doi     = {10.1038/nature10920}
}

@article{Bruder_PRL_2016,
  title = {Genuine Quantum Signatures in Synchronization of Anharmonic Self-Oscillators},
  author = {L\"orch, Niels and Amitai, Ehud and Nunnenkamp, Andreas and Bruder, Christoph},
  journal = {Phys. Rev. Lett.},
  volume = {117},
  issue = {7},
  pages = {073601},
  numpages = {6},
  year = {2016},
  month = {Aug},
  publisher = {American Physical Society},
  doi = {10.1103/PhysRevLett.117.073601}
}

@article{Dykman_Krivoglaz_1984,
title = {Fluctuations in nonlinear systems near bifurcations corresponding to the appearance of new stable states},
journal = {Physica A: Statistical Mechanics and its Applications},
volume = {104},
number = {3},
pages = {480-494},
year = {1980},
issn = {0378-4371},
doi = {https://doi.org/10.1016/0378-4371(80)90010-2},
url = {https://www.sciencedirect.com/science/article/pii/0378437180900102},
author = {M.I. Dykman and M.A. Krivoglaz}
}

@article{Liang_PRA_2014,
  title = {Symmetries and conserved quantities in Lindblad master equations},
  author = {Albert, Victor V. and Jiang, Liang},
  journal = {Phys. Rev. A},
  volume = {89},
  issue = {2},
  pages = {022118},
  numpages = {14},
  year = {2014},
  month = {Feb},
  publisher = {American Physical Society},
  doi = {10.1103/PhysRevA.89.022118}
}

@article{Buca_Prosen_NJP_2012,
  author = {Bu\v{c}a, Berislav and Prosen, Toma\v{z}},
  title = {A note on symmetry reductions of the Lindblad equation: transport in constrained open spin chains},
  journal = {New Journal of Physics},
  volume = {14},
  pages = {073007},
  year = {2012},
  doi = {10.1088/1367-2630/14/7/073007}
}

@article{ciuti_PRA_2017,
  title = {Critical dynamical properties of a first-order dissipative phase transition},
  author = {Casteels, W. and Fazio, R. and Ciuti, C.},
  journal = {Phys. Rev. A},
  volume = {95},
  issue = {1},
  pages = {012128},
  numpages = {5},
  year = {2017},
  month = {Jan},
  publisher = {American Physical Society},
  doi = {10.1103/PhysRevA.95.012128}
}

@article{Roukes_PRL_2012,
  title = {Passive Phase Noise Cancellation Scheme},
  author = {Kenig, Eyal and Cross, M. C. and Lifshitz, Ron and Karabalin, R. B. and Villanueva, L. G. and Matheny, M. H. and Roukes, M. L.},
  journal = {Phys. Rev. Lett.},
  volume = {108},
  issue = {26},
  pages = {264102},
  numpages = {5},
  year = {2012},
  month = {Jun},
  publisher = {American Physical Society},
  doi = {10.1103/PhysRevLett.108.264102}
}

@article{Esslinger_Science_2019,
author = {Nishant Dogra  and Manuele Landini  and Katrin Kroeger  and Lorenz Hruby  and Tobias Donner  and Tilman Esslinger },
title = {Dissipation-induced structural instability and chiral dynamics in a quantum gas},
journal = {Science},
volume = {366},
number = {6472},
pages = {1496-1499},
year = {2019},
doi = {10.1126/science.aaw4465}
}

@article{Leonard_Nature_2017,
  author  = {Julian L{\'e}onard and Andrea Morales and Philip Zupancic and Tilman Esslinger and Tobias Donner},
  title   = {Supersolid formation in a quantum gas breaking a continuous translational symmetry},
  journal = {Nature},
  volume  = {543},
  pages   = {87--90},
  year    = {2017},
  doi     = {10.1038/nature21067}
}

@article{Iacopo_PRA_2007,
  title = {Goldstone mode of optical parametric oscillators in planar semiconductor microcavities in the strong-coupling regime},
  author = {Wouters, Michiel and Carusotto, Iacopo},
  journal = {Phys. Rev. A},
  volume = {76},
  issue = {4},
  pages = {043807},
  numpages = {9},
  year = {2007},
  month = {Oct},
  publisher = {American Physical Society},
  doi = {10.1103/PhysRevA.76.043807}
}

@article{Graham_PRA_1970,
title = {Theory of cross-correlation of signal and idler in parametric oscillators},
journal = {Physics Letters A},
volume = {32},
number = {6},
pages = {373-374},
year = {1970},
issn = {0375-9601},
doi = {https://doi.org/10.1016/0375-9601(70)90005-8},
author = {R. Graham}
}

@article{Reid_Drummond_PRA_1989,
  title = {Correlations in nondegenerate parametric oscillation: Squeezing in the presence of phase diffusion},
  author = {Reid, M. D. and Drummond, P. D.},
  journal = {Phys. Rev. A},
  volume = {40},
  issue = {8},
  pages = {4493--4506},
  numpages = {0},
  year = {1989},
  month = {Oct},
  publisher = {American Physical Society},
  doi = {10.1103/PhysRevA.40.4493}
}

@article{Marsden_RMP_2007,
  title = {Dissipation-induced instabilities in finite dimensions},
  author = {Krechetnikov, R. and Marsden, J. E.},
  journal = {Rev. Mod. Phys.},
  volume = {79},
  issue = {2},
  pages = {519--553},
  numpages = {0},
  year = {2007},
  month = {Apr},
  publisher = {American Physical Society},
  doi = {10.1103/RevModPhys.79.519}
}

@article{Courtois_1991_OPO,
  author    = {J. Y. Courtois and A. Smith and C. Fabre and S. Reynaud},
  title     = {Phase Diffusion and Quantum Noise in the Optical Parametric Oscillator: A Semiclassical Approach},
  journal   = {Journal of Modern Optics},
  volume    = {38},
  number    = {2},
  pages     = {177--191},
  year      = {1991},
  doi       = {10.1080/09500349114550201}
}

@article{Ciuti_PRA_2018,
  title = {Spectral theory of Liouvillians for dissipative phase transitions},
  author = {Minganti, Fabrizio and Biella, Alberto and Bartolo, Nicola and Ciuti, Cristiano},
  journal = {Phys. Rev. A},
  volume = {98},
  issue = {4},
  pages = {042118},
  numpages = {13},
  year = {2018},
  month = {Oct},
  publisher = {American Physical Society},
  doi = {10.1103/PhysRevA.98.042118}
}

@article{Conti_PRA_2025,
  title = {Quantum hyperspins: Highly nonclassical collective behavior in quantum-optical parametric oscillators},
  author = {Calvanese Strinati, Marcello and Conti, Claudio},
  journal = {Phys. Rev. A},
  volume = {111},
  issue = {4},
  pages = {043712},
  numpages = {13},
  year = {2025},
  month = {Apr},
  publisher = {American Physical Society},
  doi = {10.1103/PhysRevA.111.043712}
}

@book{LandauLifshitzStatPhys,
  title = {Statistical Physics, Part 1},
  author = {Landau, L. D. and Lifshitz, E. M.},
  series = {Course of Theoretical Physics},
  volume = {5},
  year = {1980},
  publisher = {Pergamon Press},
  address = {Oxford}
}

@book{Andronov1966,
  author    = {Andronov, A. A. and Vitt, A. A. and Khaikin, S. E.},
  title     = {Theory of Oscillators},
  publisher = {Pergamon Press},
  address   = {Oxford},
  year      = {1966},
  doi       = {10.1016/C2013-0-06631-5}
}

@misc{Solanki2026,
  author = {Solanki, Parvinder and Cabot, Albert and Iemini, Fernando and Carollo, Federico and Krishna, Midhun and Ibrahim, Yeshma and Hajdu{\v{s}}ek, Michal and Lesanovsky, Igor and Fazio, Rosario and Zambrini, Roberta and Vinjanampathy, Sai},
  title  = {Quantum Synchronization},
  year          = {2026},
  eprint        = {2607.19328},
  archivePrefix = {arXiv},
  primaryClass  = {quant-ph}
}

@article{Clerk2010,
  author    = {Clerk, Aashish A. and Devoret, Michel H. and Girvin, Steven M.
               and Marquardt, Florian and Schoelkopf, Robert J.},
  title     = {Introduction to Quantum Noise, Measurement, and Amplification},
  journal   = {Reviews of Modern Physics},
  volume    = {82},
  number    = {2},
  pages     = {1155--1208},
  year      = {2010},
  month     = {apr},
  publisher = {American Physical Society},
  doi       = {10.1103/RevModPhys.82.1155}
}

\newpage 

\appendix
\onecolumngrid
\crefalias{section}{appendix}
\crefalias{subsection}{appendix}

\section{Stability Analysis of semi-classical EOM} \label{app_stability}
In this section, we show that the semiclassical solutions are stable against perturbation by performing a linear stability analysis. Recall that semi-classical equation of motion of the two-mode system is given by
\begin{align}
    \dot{a}_i = -i 2u n a_i + i 2D_i a_i^* - \frac{\kappa}{2}a_i + h \sum_{j=1}^2 J_{j}a_j 
\end{align}
where $n = |a_1|^2 + |a_2|^2$ and $J = -i\sigma_y$ is a 2 $\times $ 2 matrix. Writing $a_i = e^{i\theta}(x_i + i y_i)$, we pick $\theta = \frac{1}{2}\arcsin(\kappa/4D)$ to select the $x$-quadrature  whose damping is exactly canceled by the coherent drive, and only the $y$-quadrature are damped. We arrive at
\begin{align} \label{semi_EOM}
    \dot{x}_i &=  \left(2un + 2D \cos(2\theta) \right) y_i + h J_{ij}x_j \\
    \dot{y}_i &=  \left(-2un + 2D \cos(2\theta) \right) x_i - \kappa y_i + h J_{ij} y_j
\end{align}
We would like to go into a rotating frame to eliminate the $h$ term. In the rotating frame, we define $\Tilde{x}_i = (e^{-hJt})_{ij}x_j $ and $\Tilde{y}_i = (e^{-h J t})_{ij} y_j$. One steady-state solution is $y_i = 0$ and the $x$-quadrature has a fixed value determined by the constraint $x_1^2 + x_2^2 = n_{\rm ss} =  \frac{D}{u}\sqrt{1-\kappa^2/16D^2}$. For $D < \kappa/4$, the steady-state is trivial with $n_{\rm ss} = 0$, and the threshold transition occurs at $D = \kappa /4$. 

We linearize the solution around the steady-state solution $\tilde{\vb{x}} = \tilde{\vb{x}}_0 + \delta \tilde{\vb{x}}, \tilde{\vb{y}} = \delta \tilde{\vb{y}}$ such that $\tilde{\vb{x}}^T_0 \tilde{\vb{x}}_0 = n_{\rm ss}$, the EOM becomes
\begin{align}
    \delta \dot{\tilde{\vb{x}}} = 4un_{\rm ss} \delta \tilde{\vb{y}}, \qquad  \delta \dot{\tilde{\vb{y}}}  = -4u(\delta \tilde{\vb{x}}\cdot \tilde{\vb{x}}_0) \tilde{\vb{x}}_0 - \kappa \delta \tilde{\vb{y}}
\end{align}
Decompose the vector into radial and tangential directions to the attractor manifold via
\begin{align}
    \delta \tilde{\vb{x}} = \delta \tilde{x}_{\perp} \hat{\vb{r}}  + \delta  \tilde{\vb{x}}_{\|}, \qquad \delta \tilde{\vb{y}} =  \delta \tilde{y}_{\perp} \hat{\vb{r}}  + \delta \tilde{\vb{y}}_{\|}
\end{align}
where $\hat{\vb{r}}$ is a unit vector such that $\tilde{\vb{x}}_0 = \sqrt{n_{\rm ss}}\hat{\vb{r}}$ and $\delta \tilde{\vb{x}}_{\|} \cdot \hat{\vb{r}} = 0, \delta \tilde{\vb{y}}_{\|} \cdot \hat{\vb{r}} = 0$, we obtain
\begin{align}\label{linearized_EOM}
    \delta \dot{\tilde{\vb{x}}}_{\|} &= 4 u n_{\rm ss} \delta \tilde{\vb{y}}_{\|}, \qquad \delta \dot{\tilde{\vb{y}}}_{\|} = -\kappa \delta \tilde{\vb{y}}_{\|} \\
    \delta \dot{\tilde{x}}_{\perp} &= 4 u n_{\rm ss} \delta \tilde{y}_{\perp}, \qquad  \delta \dot{\tilde{y}}_{\perp} = -4u n_{\rm ss} \delta \tilde{x}_{\perp} - \kappa \delta \tilde{y}_{\perp}
\end{align}
The radial direction $\delta \dot{\tilde{x}}_{\perp}, \delta \dot{\tilde{y}}_{\perp}$ are both damped. 
The eigenvalue of the tangential mode for $\delta \tilde{x}_{\|}$ and $\delta \tilde{y}_{\|}$ are $\lambda = 0$, $\lambda = -\kappa$, respectively. The tangential fluctuation $\delta \tilde{\vb{x}}$ changes the direction of the radial vector $\vb{r}$ of the $S^1$ manifold and corresponds to the gapless phase modes. The above stability analysis generalizes naturally to the $N$-mode case. We find $\delta \tilde{\vb{x}}_{\|}$ to be a $N-1$ dimensional vector tangential to the $S^{N-1}$ sphere, with $N-1$ gapless modes. 

Let $\hat{\vb{r}} = (\cos \phi, \sin \phi)$. Then $\delta \tilde{\vb{x}}_{\|} = \delta \tilde{x}_{\|} \hat{\vb{s}} $ with $\hat{\vb{s}} = \partial_\phi \hat{\vb{r}} = (-\sin\phi, \cos \phi)$ and $\delta \tilde{x}_{\|}(t) = \sqrt{n_{\rm ss}} \phi(t)$. The EOM for the phase variable, in the unrotated frame, is
\begin{align}\label{two_mode_phase_EOM_app}
    \dot{\phi} = h + 4 u \sqrt{n_{ss}} \delta y_{\|}, \qquad \delta \dot{y}_{\|} = -\kappa \delta y_{\|}
\end{align}
The phase mode rotates at frequency $h$ and couples to the amplitude fluctuation of the $y$-quadrature in the tangential direction.

For $N$ modes, the unit vector is a function of $N-1$ angular variables, given by $\hat{\vb{r}} = \vb{r}(\phi^1, \cdots, \phi^{N-1})$. Decomposing the tangent vector in terms of $\phi^\alpha$ with $\alpha = 1, \cdots N - 1$ gives
\begin{align} \label{angular_comp_EOM}
    \delta \tilde{\vb{x}}_{\|} = \sqrt{n_{\rm ss}} \phi^\alpha \partial_\alpha \hat{\vb{r}}, \qquad \delta \tilde{\vb{y}}_{\|} = \delta y^\alpha \partial_\alpha \hat{\vb{r}}
\end{align}
To find the oscillation frequency in the lab frame, we first need to introduce a metric on $S^{N-1}$. Since $d\hat{\vb{r}} = \partial_\alpha \hat{\vb{r}} d\phi^\alpha$, we find the induced metric to be
\begin{align}
    ds^2 = d\hat{\vb{r}} \cdot d\hat{\vb{r}} = \partial_\alpha \hat{\vb{r}} \cdot \partial_\beta \hat{\vb{r}} d\phi^\alpha d\phi^\beta \implies g_{\alpha \beta} = \partial_\alpha \hat{\vb{r}} \cdot \partial_\beta \hat{\vb{r}}
\end{align}
In the lab frame, $hK\hat{\vb{r}}$ is the direction of the drift tangent to the manifold. 
Expanding in the basis of tangent vectors gives $hK\hat{\vb{r}} = \omega^\alpha \partial_\alpha \hat{\vb{r}}$, where we find 
\begin{align}
    \omega^\alpha = h g^{\alpha \beta} \partial_\beta \hat{\vb{r}} \cdot K \hat{\vb{r}}
\end{align}
where $g^{\alpha \beta} g_{\beta \gamma} = \delta^{\alpha}_\gamma$. Using Eq.~\eqref{linearized_EOM} and Eq.~\eqref{angular_comp_EOM}, we find the EOM for the phases to be
\begin{align}
    \dot{\phi}^\alpha = \omega^\alpha + 4u \sqrt{n_{\rm ss}} \delta y^\alpha, \qquad \delta \dot{y}^\alpha = - \kappa \delta y^\alpha
\end{align}
for $\alpha = 1, \cdots, N -1$ gapless mode. We can use the metric to lower the index and obtain $\dot{\phi}_\alpha = g_{\alpha \beta} \dot{\phi}^\beta$. 

In practice, finding the frequency $\omega^\alpha$ is complicated. To make the dynamics manifest, one should work in a coordinate system where the antisymmetric matrix $K$ is brought to a block canonical form:
$K = O^T \Sigma O$, where 
\begin{align}
    \Sigma = \bigoplus_{r=1}^{\lfloor N/2 \rfloor} \lambda_r (i\sigma_y) \oplus \begin{cases}
        0 \qquad \text{if N is odd}\\
        \emptyset \qquad \text{if N is even}
    \end{cases}
\end{align}
Immediately, one sees that there are only $\lfloor N/2 \rfloor$ independent frequencies $h\lambda_r$, instead of the naive $N-1$ frequencies. The rest of the non-oscillating gapless mode corresponds to the radial degree of freedom that sets the radius of the circle in each of the $2 \times 2$ blocks. Let $\vb{q} = O \vb{x}$ and $\vb{p} = O\vb{y}$ be the rotated quadrature. Restricting to $(2r-1, 2r)$ blocks, Eq.~\eqref{semi_EOM} becomes
\begin{align}
    \dot{\vb{q}}^{(r)} &= (2un + 2D \cos(2\theta)) \vb{p}^{(r)} + h \lambda_r J \vb{q}^{(r)} \\
    \dot{\vb{p}}^{(r)} &= (-2un + 2D \cos(2\theta)) \vb{q}^{(r)} - \kappa \vb{p}^{(r)} + h\lambda_r J \vb{p}^{(r)}
\end{align}
Let's work in the rotating frame with frequency $h \lambda_r$ for each block, which gives
\begin{align}
    \dot{\vb{q}}^{(r)} &= (2un + 2D \cos(2\theta)) \vb{p}^{(r)} \\
    \dot{\vb{p}}^{(r)} &= (-2un + 2D \cos(2\theta)) \vb{q}^{(r)} - \kappa \vb{p}^{(r)} 
\end{align}
Expanding around the steady-state solution to study fluctuations, we write
\begin{align}
    \vb{q}^{(r)}_0 = \rho_r \hat{\vb{r}}, \qquad \delta \vb{q}^{(r)} = \delta \rho_r \hat{\vb{r}} + \rho_r \theta_r \hat{\vb{s}}, \qquad 
    \delta \dot{\vb{q}}^{(r)} = 4un_{\rm ss} \delta \vb{p}^{(r)}
\end{align}
subject to the constraint $\sum_{r=1}^{\lfloor N/2 \rfloor} \rho_r^2 + \varsigma^2 = n_{\rm ss}$.
From which we find
\begin{align}
    \delta \dot{\rho}_r = 4un_{\rm ss} \delta p_{\perp,r}, \qquad \dot{\theta}_r = \frac{4un_{\rm ss}}{\rho_r} \delta p_{\|, r}
\end{align}
where we decomposed $\delta \vb{p}^{(r)} = \delta p_{\perp,r} \hat{\vb{r}} + \delta p_{\|,r} \hat{\vb{s}}$, which has EOM
\begin{align}
    \delta \dot{p}_{\perp, r} = -4u \left(\sum_{r'} \rho_{r'} \delta \rho_{r'}\right) \rho_r - \kappa \delta p_{\perp,r}, \qquad \delta \dot{p}_{\|, r} = -\kappa \delta p_{\|, r}
\end{align}
In the long-time limit, the fluctuations $\delta p_{\perp,r}, \delta p_{\|,r}$ damp away, and we are left with, in the lab frame
\begin{align}
    \dot{\rho}_r = 0, \qquad \dot{\theta}_r = h \lambda_r
\end{align}
Within each $2\times 2$ block, it picks a circle with radius $\rho_r$ with frequency $h \lambda_r$.

\section{Quantum Exact Solution} \label{app_exact}

In this section, we derive the exact steady-state density matrix by finding a purification of the original system. A priori, finding the correct purification requires knowing the exact spectrum of the original system. However, as pointed out in \cite{two_photon_exact}, the original system can be purified by finding another system that can coherently absorb all the photon loss, which in this case is the system itself. These models are solvable via the coherent quantum absorber method \cite{CQA} and exhibit ``hidden" time-reversal symmetry \cite{hTRS} in the steady-state. 

We first make some observations. The hopping term is proportional to the generator of the $O(N)$ symmetry. Since our steady-state is unique, the steady-state is invariant under the $O(N)$ action. We set $h = 0$ in the following derivation as it will not change the steady-state. A non-zero $h$ leads to a pair of imaginary eigenvalues whose real part decays to zero in the thermodynamic limit, generating metastable persistent oscillation. Consider the cascaded master equation of the doubled system
\begin{align}
    \partial_t \hat{\rho}_{LR} = -i[{\hat{H}_{LR}}, \hat{\rho}_{LR}] + \kappa \sum_{i=1}^N \mathcal{D}[\hat{a}_{i,L} - \hat{a}_{i,R}] \hat{\rho}_{LR}
\end{align}
where $\hat{H}_{LR} = \hat{H}_L - \hat{H}_R - \frac{i}{2}\kappa \sum_{i=1}^N(\hat{a}_{i,L}^\dagger \hat{a}_{i,R} - \hat{a}_{i,L} \hat{a}^\dagger_{i,R} )$ with $\hat{H}_{L} = \hat{H} \otimes \mathbb{I}, \hat{H}_{R} = \mathbb{I} \otimes \hat{H}$. For the steady-state to be pure, it must satisfy
\begin{align}
    \hat{H}_{LR} \ket{\psi} = 0, \qquad (\hat{a}_{i,L} - \hat{a}_{i,R}) \ket{\psi} = 0
\end{align}
Requiring the pure steady-state to be the dark state of the collective jump operators allows us to rewrite the constraints:
\begin{align}
    (\hat{H}_{\rm eff, L} - \hat{H}_{\rm eff, R})\ket{\psi} = 0, \qquad (\hat{a}_{i,L} - \hat{a}_{i,R}) \ket{\psi} = 0
\end{align}
where $\hat{H}_{\rm eff} = \hat{H} - \frac{i}{2}\kappa \sum_i \hat{a}_i^\dagger \hat{a}_i$. If the constraints are solvable, we can obtain an expression for $\ket{\psi}$, and the steady-state of the original system is obtained by tracing out the auxiliary system, given by $\rho_{\rm ss} = \Tr_{R} \ket{\psi}\bra{\psi}$. The pure state $\ket{\psi}$ is then the purification of the original system.

The collective jump operator constraint motivates us to define 
\begin{align}
    \hat{\alpha}_i = \frac{\hat{a}_{i,L} + \hat{a}_{i,R}}{\sqrt{2}}, \qquad \hat{\beta}_i = \frac{\hat{a}_{i,L} - \hat{a}_{i,R}}{\sqrt{2}}
\end{align}
satisfying
\begin{align}
    [\hat{\alpha}_i, \hat{\alpha}_j^\dagger] = [\hat{\beta}_i, \hat{\beta}_j^\dagger] = \delta_{i,j}, \qquad [\hat{\alpha}_i, \hat{\beta}_j] = 0
\end{align}
Using the second constraint $\hat{\beta}_i \ket{\psi} = 0$, we can rewrite the first constraint in terms of the symmetric and antisymmetric modes $\hat{\alpha}_i, \hat{\beta}_i$ as
\begin{align}
    \left(u(\hat{N}_\alpha + \hat{N}_\beta)\sum_{i=1}^N\left(\hat{\alpha}_i^\dagger \hat{\beta}_i + \hat{\beta}_i^\dagger \hat{\alpha}_i \right) - 2D\sum_{i=1}^N\left(\hat{\alpha}_i^\dagger \hat{\beta}_i^\dagger + h.c. \right) - \frac{i}{2}\kappa \sum_{i=1}^N \left(\hat{\alpha}_i^\dagger \hat{\beta}_i + \hat{\beta}_i^\dagger \hat{\alpha}_i \right) \right) \ket{\psi} &= 0 \\
    \left(u(\hat{N}_\alpha + \hat{N}_\beta)\sum_{i=1}^N \hat{\beta}_i^\dagger \hat{\alpha}_i  - 2D\sum_{i=1}^N \hat{\alpha}_i^\dagger \hat{\beta}_i^\dagger  - \frac{i}{2}\kappa \sum_{i=1}^N  \hat{\beta}_i^\dagger \hat{\alpha}_i \right) \ket{\psi} &= 0
\end{align}
where $\hat{N}_\alpha = \sum_{i=1}^N \alpha^\dagger_i \alpha_i$ and $\hat{N}_\beta = \sum_{i=1}^N \beta^\dagger_i \beta_i$. Factoring out $\hat{\beta}_i^\dagger$ and dropping $\hat{N}_\beta$ using the second constraint, we find
\begin{align}
    \sum_i \hat{\beta}_i^\dagger \left( \left(u(\hat{N}_\alpha + 1) - \frac{i}{2}\kappa\right) \hat{\alpha}_i - 2D \hat{\alpha}_i^\dagger \right) \ket{\psi} = 0
\end{align}
The steady-state is factorized into symmetric and antisymmetric part $\ket{\psi} = \ket{\psi_\alpha} \ket{\psi_\beta}$ and $\ket{\psi_\beta} = \ket{0}$ such that it is annihilated by $\hat{\beta}_i$. The symmetric part satisfies
\begin{align}
     \left(\left(u(\hat{N}_\alpha + 1) - \frac{i}{2}\kappa\right) \hat{\alpha}_i - 2D \hat{\alpha}_i^\dagger \right) \ket{\psi_\alpha} = 0
\end{align}
Define pair-creation operator $\hat{K}_+ = \frac{1}{2}\sum_{i=1}^N (\hat{\alpha}_i^\dagger)^2$. This has the following nice commutation relation
\begin{align}
    [\hat{\alpha}_i, \hat{K}_+] = \hat{\alpha}_i^\dagger, \qquad [\hat{\alpha}_i^\dagger, \hat{K}_+] = 0, \qquad 
    [\hat{N}_\alpha, \hat{K}_+] = 2 \hat{K}_+ 
\end{align}
One can also show that $[\hat{N}_\alpha, (\hat{K}_+)^m] = 2m (\hat{K}_+)^m$. Using the ansatz 
\begin{align}
    \ket{\psi_\alpha} = \sum_{m=0}^\infty c_m (\hat{K}_+)^m \ket{0}
\end{align}
gives the following recursion relation:
\begin{align}
    m\left(2m u - \frac{i}{2} \kappa\right) c_m - 2D c_{m-1} = 0, \qquad \text{for $m \geq 1$}
\end{align}
which allows us to solve for the coefficient $c_m$: 
\begin{align}
    c_{m} = \left(\frac{D}{u} \right)^{m} \frac{1}{m! (1-\frac{i\kappa}{4u})_m} c_0, \qquad c_0 = 1
\end{align}
Therefore, the (un-normalized) steady-state takes the form
\begin{align}
    \ket{\psi} = \sum_{m=0}^\infty \frac{1}{m! (\delta)_m} \left( \frac{D}{u}\right)^m \left(\frac{1}{2}\sum_{i=1}^N \hat{\alpha}_i^\dagger \hat{\alpha}_i^\dagger \right)^m \ket{0}_\alpha \ket{0}_\beta 
\end{align}
with $\delta = 1 - \frac{i\kappa}{4u}$. The normalization is given by
\begin{align}
    \bra{\psi}\ket{\psi} &= \sum_{m=0}^\infty \frac{1}{m! (m')!(\delta)_m (\delta^*)_{m'}} \left(\frac{D}{u} \right)^{m} \left(\frac{D}{u} \right)^{m'} \mel{0}{(\hat{K}_-)^{m'} (\hat{K}_+)^m}{0} \\
    &= \sum_{m=0}^\infty \frac{(N/2)_m}{(\delta)_m (\delta^*)_m} \frac{1}{m!} \left(\frac{D}{u}\right)^{2m} \\
    &= {}_1 F_2(N/2; \delta, \delta^*; (D/u)^2) 
\end{align}
In the second line, we used generalized Binomial expansion to write
\begin{align}
    (\hat{K}_+)^m = \frac{1}{2^m}\sum_{k_1 \cdots k_N = m} \frac{m!}{k_1! \cdots k_N!} \prod_{i=1}^{N} (\alpha_i^\dagger)^{2k_i}
\end{align}
which gives
\begin{align}
    \mel{0}{(\hat{K}_-)^{m'} (\hat{K}_+)^{m}}{0} = \delta_{m,m'} \frac{1}{2^m 2^m} \sum_{\sum_i^N k_i = m} \frac{(m!)^2}{(k_1! \cdots k_N!)^2} \prod_{i=1}^N (2k_i)! = (m!)(N/2)_m
\end{align}
For two-mode model, we have $\bra{\psi}\ket{\psi} = {}_1 F_2(1; \delta, \delta^*; (D/u)^2) $.

\section{Mean photon number and Fano ratio from Exact Solution} \label{app_obs}
In this section, we will use the exact solution to compute some observables. Let $\lambda = (D/u)^2$. We define a generating function
\begin{align}
    Z(\lambda) \equiv \bra{\psi}\ket{\psi} = {}_1 F_2(N/2; \delta, \delta^*; \lambda)
\end{align}
To compute the mean photon number, we first observe that $\hat{N}(\hat{K}_+)^m\ket{0} = 2m (\hat{K}_+)^m \ket{0}$. Thus, $(\hat{K}_+)^m\ket{0}$ contains $2m$ excitations, distributed equally among the left and right mode. The unnormalized expectation value of $\hat{N} = \sum_{i=1}^N \alpha_i^\dagger \alpha_i$ is
\begin{align}
    \mel{\psi}{\hat{N}}{\psi} = \sum_{m=0}^\infty (2m)\frac{(N/2)_m}{(\delta)_m (\delta^*)_m} \frac{1}{m!}\left(\frac{D}{u} \right)^{2m} = 2 \lambda Z'(\lambda) 
\end{align}
For the convenience of calculation, we introduce a new set of state defined by $\ket{m} \equiv (K_+)^m\ket{0}$, which tracks the number of pair excitations. Note that this set of state $\{\ket{m}\}$ are orthogonal but not complete. We also introduce an operator $\hat{m}$ that probe the number of pair excitation such that $\hat{m}(\hat{K}_+)^m\ket{0} = m (\hat{K}_+)^m\ket{0}$. 
We find that 
\begin{align}
    \langle \hat{m} \rangle = \frac{1}{2}\frac{\mel{\psi}{\hat{N}}{\psi}}{\bra{\psi}\ket{\psi}} = \frac{\lambda Z'(\lambda)}{Z(\lambda)}
\end{align}

The mean photon number of the original system $\hat{n} = \sum_{i=1}^N \hat{a}_{i,L}^\dagger \hat{a}_{i,L}$ in terms of $\hat{\alpha}_i, \hat{\beta}_i$ are given by
\begin{align}
    \hat{n} = \frac{1}{2}\sum_{i=1}^N \left(\hat{\alpha}^\dagger_{i}\hat{\alpha}_{i} + \hat{\beta}_i^\dagger \hat{\alpha}_i + \hat{\alpha}_i^\dagger \hat{\beta}_i + \hat{\beta}_i^\dagger \hat{\beta}_i \right)
\end{align}
To compute the mean photon number, we first note that our steady-state are of the form $\ket{\psi} = \sum_{m}c_m \ket{m}$, and the relevant matrix elements are
\begin{align}
    \mel{m'}{\hat{n}}{m} = \delta_{m,m'} m, \qquad \mel{m'}{\hat{n}^2}{m} = \delta_{m,m'} \left( m^2 + \frac{m}{2} \right)
\end{align}
It follows that
\begin{align}
    \langle \hat{n} \rangle = \langle \hat{m} \rangle = \lambda \frac{(N/2)}{\delta \delta^*}\frac{{}_1 F_2(N/2+1; \delta+1, \delta^*+1; \lambda)}{{}_1 F_2(N/2; \delta, \delta^*; \lambda)}
\end{align}
where we note that $\frac{d}{d\lambda}{}_1 F_{2}(a; b, c; \lambda) = \frac{a}{bc}{}_1 F_{2}(a+1; b+1, c+1; \lambda)$. Furthermore, from
\begin{align}
    \langle \hat{n}^2 \rangle = \langle \hat{m}^2 \rangle + \frac{\langle \hat{m} \rangle}{2}
\end{align}
we find the variance of $\hat{n}$ is given by 
\begin{align}
    \text{Var}(\hat{n}) = \langle \hat{n}^2 \rangle - \langle \hat{n} \rangle^2 = \langle \hat{m}^2 \rangle + \frac{\langle \hat{m} \rangle}{2} - \langle \hat{m} \rangle^2 =  \text{Var}(\hat{m}) + \frac{1}{2} \langle \hat{m} \rangle
\end{align}
Using $\langle \hat{m}(\hat{m}-1) \rangle = \lambda^2 \frac{Z''(\lambda)}{Z(\lambda)}$, we find
\begin{align}
    \text{Var}(\hat{m}) = \lambda^2 \frac{(N/2)(N/2+1)}{\delta \delta^* (\delta+1)(\delta^*+1)} \frac{{}_1 F_{2}(N/2+2; \delta+2, \delta^*+2; \lambda)}{{}_1 F_{2}(N/2; \delta, \delta^*; \lambda)}  + \langle m \rangle - \langle m \rangle^2
\end{align}
The Fano ratio, computed from the exact solution, is given by
\begin{align} \label{exact_fano}
    \text{Fano} = \frac{\text{Var}(\hat{n})}{\langle \hat{n} \rangle} = \frac{\text{Var}(\hat{m})}{\langle \hat{m} \rangle} + \frac{1}{2} 
\end{align}
In the limit of $\kappa \rightarrow 0$, $\frac{\text{Var}(\hat{m})}{\langle \hat{m} \rangle} \rightarrow \frac{1}{2}$, and the steady-state is Poisson distributed.

It would also be of interest to compute the Fano ratio from the semiclassical EOM to see when the semiclassical approximation matches the quantum solution. Surprisingly, it matches well with the exact solution down to small photon number. We will try to compute the mean and variance of $n$ from the linearized semiclassical EOM. We note that since $n = |\vb{x}|^2 + |\vb{y}|^2$, we can expand to first order in fluctuation and find
\begin{align}
    n = x_0^2 + 2 x_0 \delta x_{\perp}
\end{align}
where we recall that $\delta x_{\perp}$ is fluctuations normal to the attractor manifold $S^{N-1}$. Since fluctuations have zero mean, this implies
\begin{align}
    \langle n \rangle = x_0^2, \qquad \langle n^2 \rangle = x_0^4 + 4 x_0^2 \langle \delta x_{\perp}^2 \rangle
\end{align}
Therefore, the Fano ratio is given by
\begin{align}
    \text{Fano} = 4 \langle \delta x_{\perp}^2 \rangle
\end{align}
Recall from the linearized EOM that
\begin{align}
    \partial_t \begin{pmatrix}
        \delta x_{\perp} \\
        \delta y_{\perp}
    \end{pmatrix} = 
    \begin{pmatrix}
        0 & 4 u n_{\rm ss} \\
        -4un_{\rm ss} & -\kappa
    \end{pmatrix} 
    \begin{pmatrix}
        \delta x_{\perp} \\
        \delta y_{\perp}
    \end{pmatrix}
    + \sqrt{\kappa} 
    \begin{pmatrix}
        \xi_1 \\
        \xi_2
    \end{pmatrix}
\end{align}
with $\langle \xi_i(t) \xi_j(t') \rangle = \frac{1}{4}\delta_{ij}\delta(t-t')$. The covariance matrix satisfies the Lyapunov equation
\begin{align}
    \dot{C}(t) = A C(t) + C(t) A^T + Q
\end{align}
and at stationary, we have $0 = A C + C A^T + Q$ and gives
\begin{align}
\begin{pmatrix}
        0 & 4 u n_{\rm ss} \\
        -4un_{\rm ss} & -\kappa
    \end{pmatrix} 
    \begin{pmatrix}
        C_{11} & C_{12} \\
        C_{21} & C_{22}
    \end{pmatrix} + 
    \begin{pmatrix}
        C_{11} & C_{12} \\
        C_{21} & C_{22}
    \end{pmatrix}
    \begin{pmatrix}
        0 & -4 u n_{\rm ss} \\
        4un_{\rm ss} & -\kappa
    \end{pmatrix} 
    + \frac{1}{4}\begin{pmatrix}
        \kappa & 0 \\
        0 & \kappa
    \end{pmatrix}  = \vb{0}
\end{align}
The variance in $\delta x_{\perp}$ is given by
\begin{align}
    C_{11} = \frac{1}{4} \left( 1 + \frac{\kappa^2}{32 u^2 n_{\rm ss}^2} \right)
\end{align}
Therefore, the semiclassical calculation gives 
\begin{align}
    \text{Fano} = 1 + \frac{\kappa^2}{32 u^2 n_{\rm ss}^2} = 1+ \frac{\kappa^2}{32D^2\left(1 - \frac{\kappa^2}{16D^2} \right)} = 1 + \frac{1}{2}\frac{1}{(4D/\kappa)^2 - 1}
\end{align}
This agrees well with the quantum solution of Eq.~\eqref{exact_fano}, down to $\mathcal{O}(1)$ total photon number. Note that the semiclassical formula is only valid in the limit cycle phase, when $D > \kappa /4$ and diverges at the threshold.  

\section{Thermodynamic Limit of our Bosonic Model} \label{app_TD_limit}
In this section, we offer a complementary perspective on the thermodynamic limit used in the main text. We introduce a dimensionless parameter $n_{\rm ss}$, associated with the total photon number, and define the thermodynamic limit as $n_{\rm ss} \rightarrow \infty$. In general, a large photon number limit can be reached by either sending $u \rightarrow 0$, or $D \rightarrow \infty$, or $u \rightarrow 0$ and $ D \rightarrow \infty$ while keeping their ratio fixed. The appropriate scaling is one in which the semiclassical equation of motion, written in terms of rescaled amplitude, become independent of $n_{\rm ss}$ \cite{ciuti_PRA_2017}. This ensures that the equation of motion is well-defined, while the total photon number diverges.

For the present model, the semiclassical equation of motions is (see Eq.~\eqref{semi_coherent_eq}):
\begin{align}
     \dot{a}_i = -i 2u n a_i + i 2D a_i^* - \frac{\kappa}{2}a_i + h \sum_{j=1}^2 J_{ij}a_j
\end{align}
where $n = |a_1|^2 + |a_2|^2$. We define the scaled amplitude $\tilde{a}_i = a_i/\sqrt{n_{\rm ss}}$. With this scaling, total photon number becomes proportional to $n_{\rm ss}$ while $\tilde{a}_i \sim \mathcal{O}(1)$. The equation of motion of the scaled amplitude is
\begin{align}
    \dot{\tilde{a}}_i = -i2un_{\rm ss} |\tilde{a}|^2 \tilde{a}_i + i2D \tilde{a}_i^* -\frac{\kappa}{2} \tilde{a}_i + h \sum_{j=1}^2 J_{ij}\tilde{a}_j
\end{align}
To ensure that the equation of motion is independent of $n_{\rm ss}$, we need to hold $\tilde{u} \equiv u n_{\rm ss}$, drive $D$, and dissipation $\kappa$ fixed. In the thermodynamic limit of $n_{\rm ss} \rightarrow \infty$, the original variable scales as
\begin{align}
    u \rightarrow 0, \qquad D,\kappa \quad \text{fixed}
\end{align}
This is the same scaling as we considered in the main text.

It is useful to contrast this scaling with a bosonic mode subject to linear drive and Kerr nonlinearity. The semiclassical equation of motions is given by
\begin{align}
    i\dot{a} = (\Delta + u |a|^2) a - i\frac{\kappa}{2} a + F
\end{align}
where $\Delta$ is detuning and $F$ is linear drive strength. Following the above procedure by introducing a control parameter $n_{\rm ss}$ and a rescaled amplitude $\tilde{a} = a/\sqrt{n_{\rm ss}}$, the scaled semiclassical equation of motion becomes
\begin{align}
    i\dot{\tilde{a}} = (\Delta + u n_{\rm ss} |\tilde{a}|^2) \tilde{a}  - i\frac{\kappa}{2} \tilde{a} + \frac{F}{\sqrt{n_{\rm ss}}}
\end{align}
Ensuring the equation of motion to be independent of $n_{\rm ss}$ requires us to define
\begin{align}
    \tilde{u} = un_{\rm ss}, \qquad \tilde{F} = \frac{F}{\sqrt{n_{\rm ss}}}
\end{align}
Therefore, the thermodynamic limit $n_{\rm ss} \rightarrow \infty$ is realized by $u \rightarrow 0$, $F \rightarrow \infty$ such that their ratio $u F^2 \sim \mathcal{O}(1)$, consistent with \cite{ciuti_PRA_2017}. Unlike the parametrically driven case, a linear drive requires the driving strength to scale with $n_{\rm ss}$ in order to produce a well-defined thermodynamic limit.

\section{Steady-state Entanglement}\label{app_b_entanglement}
In this section, we review the derivation of log-negativity $E_N$ of a two-mode squeezed thermal state and discuss the behavior of $E_N$ for the $b$-mode partition described in Eq.~\eqref{b_mode_eq}. A thermal state density matrix is given by
\begin{align}
   \hat{\rho}_{\rm th} = \sum_{k=0}^\infty \frac{n_{\rm th}^k}{(n_{\rm th}+1)^{k+1}} \ket{k}\bra{k}
\end{align}
where $n_{\rm th}$ is the mean photon number. A two-mode squeezed thermal state (TMST) is obtained by applying the squeezing operator $\hat{S}(r) = e^{r(\hat{a}^\dagger \hat{b}^\dagger - \hat{a} \hat{b})}$ to the thermal density matrix
\begin{align}
    \hat{\rho}_{\rm TMST} = \hat{S}(r)[\hat{\rho}_{\rm th,A} \otimes \hat{\rho}_{\rm th,B}] \hat{S}^\dagger(r)
\end{align}
where $r$ is the squeezing strength. Since unitary evolution will not change the purity of this state, the purity is given by 
\begin{align}
    \mu = \tr(\hat{\rho}^2_{\rm TMST}) = \tr(\hat{\rho}^2_{\rm th,A}) \tr(\hat{\rho}^2_{\rm th,B}) = \frac{1}{2n_a+1} \frac{1}{2n_b+1} = \frac{1}{(2n_{\rm th}+1)^2}
\end{align}
where we assume $n_a = n_b = n_{\rm th}$. Let $\hat{N} = \hat{a}^\dagger \hat{a} + \hat{b}^\dagger \hat{b}$. The total photon number is given by $\langle \hat{N} \rangle = \tr(\hat{N} \hat{\rho}_{\rm TMST})$. Using 
\begin{align}
    \hat{S}^\dagger(r) \hat{a} \hat{S}(r) = \hat{a} \cosh(r) + \hat{b}^\dagger \sinh(r) , \qquad \hat{S}^\dagger(r) \hat{b} \hat{S}(r) = \hat{b} \cosh(r) + \hat{a}^\dagger \sinh(r)
\end{align}
we find 
\begin{align}
    \langle \hat{N} \rangle = \tr(\hat{N} \rho_{\rm TMST}) = \cosh(2r)2n_{\rm th} + 2 \sinh^2(r) = \frac{1}{\sqrt{\mu}}\cosh(2r) - 1
\end{align}
where we used $2\sinh^2(r) = \cosh(2r) - 1$. The ratio of the total photon number to the thermal photon is
\begin{align}
    \frac{\langle \hat{N} \rangle}{N_{\rm th}} = \cosh(2r) + \frac{2\sinh^2(r)}{N_{\rm th}}
\end{align}
with $N_{\rm th} = 2 n_{\rm th}$.

To compute log-negativity using the covariance matrix, we first note that the covariance matrix of the thermal state is given by $V = (n_{\rm th} + 1/2) I_4$. The squeezing operator is a Gaussian unitary, which transforms the quadrature $\vb{x} = (x_1, p_1, x_2, p_2)^T$ according to $\vb{x} \rightarrow S(r) \vb{x}$ with matrix element 
\begin{align}
    S(r) = \begin{pmatrix}
        \cosh(r) I \qquad \sinh(r) Z \\
        \sinh(r) Z \qquad \cosh(r) I
    \end{pmatrix}
\end{align}
The covariance matrix transforms as $V \rightarrow \tilde{V} = S(r) V S(r)^T$. Computing the partial transpose amounts to sending $p_2 \rightarrow -p_2$, and one finds the symplectic eigenvalue of $\tilde{V}$ to be
\begin{align}
    \tilde{\nu}_- = (n_{\rm th} + 1/2) e^{-2r}
\end{align}
Therefore, the log negativity is 
\begin{align}
    E_N = \max\{0,  -\log_2(2\tilde{\nu}_-) \} = \max\{0, \frac{2r - \ln(2n_{\rm th}+1)}{\ln(2)}\}
\end{align}
Increasing $n_{\rm th}$ leads to a decrease in $E_N$. Furthermore, we have the relation $2n_{\rm th} =  \frac{1}{\sqrt{\mu}} -1$, which allows us to write $\tilde{\nu}_- $ in terms of purity. To use TMST as a benchmark for $E_N$, we compared $E_N$ of the two models with the same total photon number and purity.

\begin{figure}
    \centering
    \includegraphics[width=0.8\linewidth]{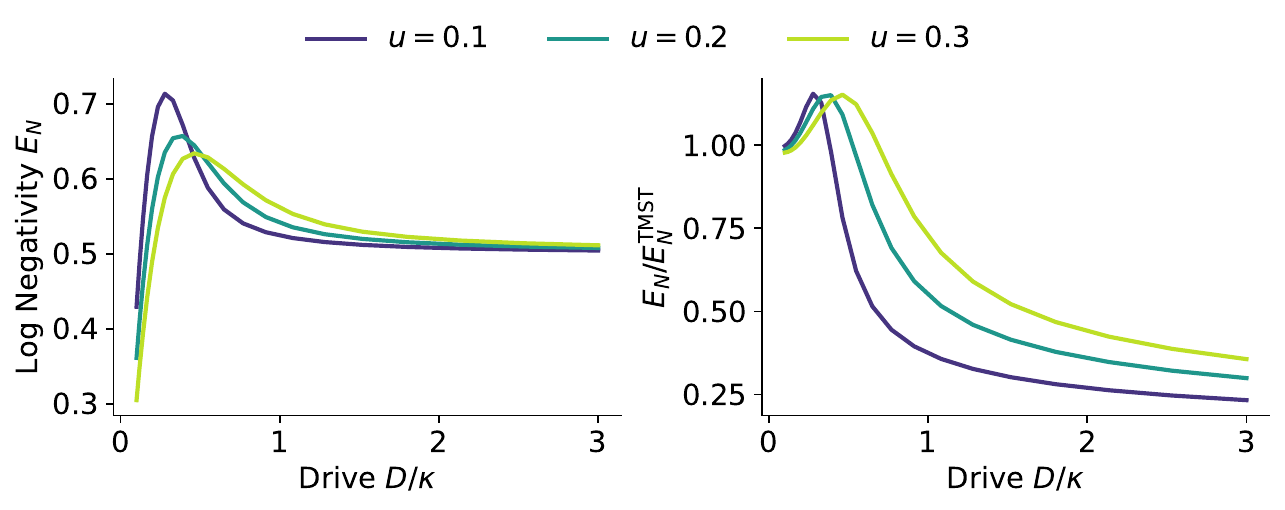}
    \caption{Log-negativity $E_N$ and the ratio of $E_N$ to that of a two-mode squeezed thermal state of the same purity and photon number in the $\hat{b}$-mode partition, where the Hamiltonian has the form of a NDPO. We set $\kappa = 1$.}
    \label{fig:b_mode_EN}
\end{figure}

Since the steady-state of our model is non-Gaussian, one needs to explicitly construct the density matrix and perform exact diagonalization of the partial transposed density matrix to obtain log-negativity. The most efficient way is to directly construct the reduced density matrix without explicitly building the creation operators. We outline the method below, which allows us to construct density matrix up to a Hilbert space truncation of $N_{\rm max} = 200$ levels per mode. The main bottleneck is thus the time required to diagonalize the matrix, and in practice we use $N_{\rm max} = 100$ level per mode. To construct the density matrix, one can utilize the fact that the steady-state only depends on two symmetric mode $\alpha_1, \alpha_2$. One can rewrite the state via
\begin{align}
    \ket{\psi} = \sum_{i=0}^{N_{\rm max}} \sum_{j=0}^{N_{\rm max}-i} A_{i,j} \ket{2i}_{s}\otimes \ket{2j}_{s} \otimes \ket{0,0}_{a}, \qquad A_{i,j} = \frac{1}{(\delta)_{i+j}}\left(\frac{D}{2u}\right)^{i+j} \frac{\sqrt{(2i)!(2j)!}}{i! j!}
\end{align}
and we can convert the symmetric/antisymmetric mode basis to left/right basis via
\begin{align}
    \ket{n}_s\ket{0}_a = \sum_{k=0}^n \frac{\sqrt{\begin{pmatrix}
        n \\
        k
    \end{pmatrix}}}{2^{n/2}} \ket{k}_L\ket{n-k}_R
\end{align}
Therefore, the wavefunction is given explicitly by
\begin{align}
    \ket{\psi} &= \sum_{i=0}^{N_{\rm max}} \sum_{j=0}^{N_{\rm max}-i} \sum_{k=0}^{2i} \sum_{p=0}^{2j} A_{i,j} \frac{\sqrt{\begin{pmatrix}
        2i \\
        k
    \end{pmatrix}}}{2^{i}} \frac{\sqrt{\begin{pmatrix}
        2j \\
        p
    \end{pmatrix}}}{2^{j}} \ket{k}_{1L} \ket{2i-k}_{1R} \ket{p}_{2L} \ket{2j-p}_{2R} \\
    &= \sum_{i=0}^{N_{\rm max}} \sum_{j=0}^{N_{\rm max}-i} \sum_{k=0}^{2i} \sum_{p=0}^{2j} A_{i,j} \frac{\sqrt{\begin{pmatrix}
        2i \\
        k
    \end{pmatrix}}}{2^{i}} \frac{\sqrt{\begin{pmatrix}
        2j \\
        p
    \end{pmatrix}}}{2^{j}} \ket{k(N_{\rm max}+1) + p}_{L} \ket{(2i-k)(N_{\rm max}+1) + (2j-p)}_{R}
\end{align}
where in the last line we reshaped $(n_1, n_2)$ into a single index $n_1 (N_{\rm max}+1) + n_2$. Schematically, let $\ket{\psi} = \sum_{\alpha, \beta} c_{\alpha, \beta} \ket{\alpha}_L \ket{\beta}_R$. Then, 
\begin{align}
    \rho &= \Tr_R \ket{\psi}\bra{\psi} = \sum_{\alpha, \alpha'}\sum_{\beta} c_{\alpha, \beta} c^*_{\alpha', \beta} \ket{\alpha}\bra{\alpha'} \\
    \implies \rho_{\alpha, \alpha'} &= \sum_{\beta} c_{\alpha, \beta} c^*_{\alpha', \beta} \qquad \text{or} \qquad  \rho = \sum_{\beta} \vb{v}_\beta \vb{v}^\dagger_\beta 
\end{align}
We directly construct the matrix element $\rho_{\alpha, \alpha'}$, which has size $(N_{\rm max}+1)^2 \cross (N_{\rm max}+1)^2$, allowing efficient construction of the density matrix.

To construct the density matrix in the $(b_1, b_2)$ basis, we used the fact that
\begin{align}
    \alpha_1 = \frac{1}{2}(b_{1L} + b_{1R} + b_{2L} + b_{2R}) , \qquad \alpha_2 = \frac{1}{2i}(b_{1L} + b_{1R} - b_{2L} - b_{2R})
\end{align}
Then, we find
\begin{align}
    \ket{n,m}_{\alpha}\ket{0,0}_\beta = \frac{i^{m} \sqrt{n! m!}}{2^{n+m}} \sum_{\sum_i^4 k_i = n}\sum_{\sum_i^4 p_i = m} (-1)^{p_3 + p_4} \prod_{i=1}^4 \frac{\sqrt{(k_i + p_i)!}}{k_i! p_i!} \ket{k_1 + p_1}_{1L}\ket{k_2 + p_2}_{1R}\ket{k_3 + p_3}_{2L}\ket{k_4 + p_4}_{2R}
\end{align}

Fig. \ref{fig:b_mode_EN} shows the log-negativity as a function of drive for various $u$ in the b-mode partition. Below threshold $(D < \kappa/4)$, we expect nonlinearity to play a small role and the Hamiltonian is that of parametric amplifier, which realizes a thermal squeezed state in the steady-state. Indeed, comparison against TMST shows a comparable value below threshold. Above threshold, the log negativity decreases towards a constant value and the ratio with TMST decreases towards zero.

\section{Symmetry-breaking Nonlinearities}\label{app_symm_break}
In this section, we show how  the two-mode model without $U(1)$ symmetry can nevertheless display limit cycle behavior over some parameter regime. The key observation is that such model can be viewed as weakly symmetry-broken version of the $U(1)$ symmetric parent model studied in the main text (see Eq.~\eqref{two_mode_model}). In a suitable rotating frame, symmetry-breaking term become fast-oscillating and can be neglected within a rotating-wave approximation (RWA), leading to effective $U(1)$ dynamics. Beyond this approximation, the symmetry-breaking terms generate phase locking, which destroys the limit cycle at sufficiently large drive. 

 The model considered in Eq.~\eqref{two_mode_model} has a weak $U(1)$ symmetry given by
\begin{align}
    \begin{pmatrix}
        \hat{a}_1 \\
        \hat{a}_2
    \end{pmatrix} \longrightarrow 
    \begin{pmatrix}
        \cos \theta & \sin \theta \\
        -\sin \theta & \cos \theta
    \end{pmatrix}
     \begin{pmatrix}
        \hat{a}_1 \\
        \hat{a}_2
    \end{pmatrix}
\end{align}
The nonlinearity respects this $U(1)$ symmetry by having an equal magnitude of self Kerr and cross Kerr. In many experiments, these couplings are not exactly equal. We therefore consider the more general interaction
\begin{align}
    \hat{H}_{\rm NL} = u_1 (\hat{n}_1^2 + \hat{n}_2^2) + 2 u_2 \hat{n}_1 \hat{n}_2 = u_1 \hat{N}^2 - 2(u_1 - u_2) \hat{n}_1 \hat{n}_2
\end{align}
where $\hat{N} = \hat{n}_1 + \hat{n}_2$, and we recover the interaction in Eq.~\eqref{two_mode_model} when $u_1 = u_2 = u$. It turns out to be easier to work in the rotated basis $\hat{b}_1, \hat{b}_2$, defined as
\begin{align}
    \hat{b}_1 = \frac{\hat{a}_1 + i \hat{a}_2}{\sqrt{2}}, \qquad \hat{b}_2 = \frac{\hat{a}_1 - i\hat{a}_2}{\sqrt{2}},
\end{align}
where the $U(1)$ symmetry acts as $\hat{b}_1 \rightarrow e^{i\theta}\hat{b}_1$ and $\hat{b}_2 \rightarrow e^{-i\theta}\hat{b}_2$. The generator of the symmetry is $\hat{Q} = \hat{b}^\dagger_1 \hat{b}_1 - \hat{b}^\dagger_2 \hat{b}_2$. In this basis, the Hamiltonian is $\hat{H} = \hat{H}_{\rm NL} + \hat{H}_{\rm drive} + \hat{H}_{\rm rotate}$, where
\begin{align}
    \hat{H}_{\rm NL} &= \frac{u_1 + u_2}{2} \hat{N}^2 + \frac{u_1 - u_2}{2}(\hat{b}_1^\dagger \hat{b}_2 + \hat{b}_2^\dagger \hat{b}_1)^2  \\
    \hat{H}_{\rm drive} &= - 2D\left( \hat{b}_1^\dagger \hat{b}_2^\dagger + \hat{b}_1 \hat{b}_2 \right) \\
    \hat{H}_{\rm rotate} &= -h (\hat{b}_1^\dagger \hat{b}_1 - \hat{b}_2^\dagger \hat{b}_2) 
\end{align}
Let $\delta u = u_1 - u_2$. We note that the term proportional to $\delta u$ does not commute with $\hat{Q}$. In the rotating frame generated by the unitary $\hat{U}(t) = e^{-ih\hat{Q}t}$, the non-commuting terms pick up an oscillatory factor $e^{\pm 4iht}$, and the nonlinear Hamiltonian becomes
\begin{align}
    \hat{H'}_{\rm NL} =  \frac{u_1 + u_2}{2} \hat{N}^2 + \frac{u_1 - u_2}{2} \left( e^{-4iht}(\hat{b}^\dagger_1 \hat{b}_2)^2 + e^{4iht}(\hat{b}_2^\dagger \hat{b}_1)^2 + 2\hat{n}_1 \hat{n}_2 + \hat{N} \right)
\end{align}
where $\hat{n}_i = \hat{b}^\dagger_i \hat{b}_i$. When $4h \gg |u_1 - u_2|$, the oscillatory term can be neglected within a RWA approximation, yielding an effective Hamiltonian with $U(1)$ symmetry:
\begin{align}
    \hat{H}_{\rm RWA} =  \frac{u_1 + u_2}{2} \hat{N}^2 + \frac{u_1 - u_2}{2}(2\hat{n}_1 \hat{n}_2 + \hat{N}) - 2D\left( \hat{b}_1^\dagger \hat{b}_2^\dagger + \hat{b}_1 \hat{b}_2 \right), \qquad [\hat{H}_{\rm RWA}, Q] = 0
\end{align}
Although the starting Hamiltonian has no $U(1)$ symmetry, the effective dynamics over an intermediate time is governed by an effective $U(1)$ symmetric Hamiltonian. Thus, we expect limit cycle behavior to exist over a range of parameters.

Since the exact dynamics is not $U(1)$ symmetric, we would like to find out how the presence of symmetry-breaking term affect the dynamics. We treat $\delta u$ to be small and derive corrections to the semiclassical equation of motion given in Eq.~\eqref{linearized_EOM}. The equation of motion of the phase variable, given in Eq.~\eqref{two_mode_phase_EOM_app}, shows that, to leading order, only the fluctuations of the $y$-quadrature tangential to the limit cycle affects the phase dynamics. Therefore, we focus only on $y$-quadrature and find that the $y$-quadrature equation of motion receives the following correction
\begin{align}
    \dot{y}_1 = 2 \, \delta u \, n_2 x_1 + \cdots, \qquad \dot{y}_2 = 2 \, \delta u \, n_1 x_2 + \cdots
\end{align}
Expanding near the semiclassical limit cycle solution $\vb{x}_0 = \sqrt{n_{\rm ss}}\hat{\vb{r}}$, and projecting to onto the tangential direction $\hat{\vb{s}} = (-\sin(\phi), \cos(\phi))$ gives the following expression of $\delta y_{\|}$:
\begin{align}
    \delta \dot{y}_{\|} = -\kappa \delta y_{\|} + \frac{\delta u}{2} n_{\rm ss}^{3/2} \sin(4\phi)
\end{align}

In the large $\kappa$ limit, we approximate $\delta y_{\|} \approx \frac{\delta u}{2\kappa}n_{\rm ss}^{3/2}\sin(4\phi)$. The equation of motion of the phase variable becomes
\begin{align}
    \dot{\phi} = h  + \frac{2 u \, \delta u \,}{\kappa} n_{\rm ss}^2 \sin(4\phi)
\end{align}
This has the form of Adler equation: the first term drives the uniform phase winding, while the second term tends to lock the phase. From this perspective, parametrically driven oscillator with persistent oscillation without $U(1)$ symmetry can be viewed as symmetry-broken version of the $U(1)$ symmetric model studied here. Related persistent beating regime in coupled parametric oscillators were studied in Ref. \cite{bello_persistent_2019}. Persistent oscillation survives only when the winding term dominates. For 
\begin{align}
    |h| < \left|\frac{2 u \, \delta u \, }{\kappa} n_{\rm ss}^2 \right|,
\end{align}
a fixed point exists and the limit cycles is replaced by a phase-locked steady-state. In the large drive $D$ limit, we have $n_{\rm ss} \propto D/u$, and the phase locking term grows as $D^2$. The boundary between phase locking and limit cycles is given by
\begin{align}
    h \propto \frac{\delta u }{u \kappa}D^2
\end{align}
For any fixed symmetry-breaking nonlinearity $\delta u$, increasing drive eventually drives the system into a phase locked regime. This suggests that the loss of persistent oscillation at strong parametric drive studied in Ref. \cite{bello_persistent_2019} can be understood as phase-locking induced by explicit symmetry-breaking.

\section{Quantum van der Pol oscillator}\label{app_van}

A $U(1)$ symmetric limit cycles model provides a clean starting point for studying synchronization and may also serve as a useful model of a laser, where a neutral phase degree of freedom is desired. A standard example is the quantum van der Pol (vdP) oscillator, in which incoherent single photon gain is balanced by a nonlinear two-photon loss in the steady-state. In this section, we highlight some interesting features of quantum vdP oscillators. In particular, we point out that the Fano ratio approaches $F=1.5$ in the thermodynamic limit, compared to our model which gives $F=1$ in the large drive (large photon) limit. 

Consider a harmonic oscillator $H = \omega a^\dagger a$ subject to linear pump and nonlinear two-photon loss. In the rotating frame with frequency $\omega$, the Lindblad master equation of the quantum vdP is
\begin{align}
    \dot{\hat{\rho}} = \kappa \mathcal{D}[\hat{a}] \hat{\rho} + \gamma_1 \mathcal{D}[\hat{a}^\dagger] \hat{\rho} + \gamma_2 \mathcal{D}[\hat{a}^2] \hat{\rho}
\end{align}
Expanding in Fock basis $\hat{\rho} = \sum_{m,n}\rho_{m,n}\ket{m}\bra{n}$, the EOM becomes
\begin{align}
    \dot{\rho}_{m,n} &= \kappa \sqrt{(m+1)(n+1)}\rho_{m+1, n+1} + \gamma_1 \sqrt{mn} \rho_{m-1, n-1} + \gamma_2 \sqrt{(m+1)(m+2)(n+1)(n+2)} \rho_{m+2, n+2} \\
    &- \left(\kappa \frac{m+n}{2}  + \gamma_1 \frac{m+1+n+1}{2} + \gamma_2 \frac{m(m-1) + n(n-1)}{2} \right)\rho_{m,n}
\end{align}
The off-diagonal component decays away, and the steady-state is diagonal in the Fock basis. This gives
\begin{align}
    \dot{\rho}_m = \kappa(m+1) \rho_{m+1} + \gamma_1 m \rho_{m-1} + \gamma_2(m+1)(m+2) \rho_{m+2} - (\kappa m + \gamma_1(m+1) + \gamma_2 m (m-1)) \rho_m
\end{align}
subject to the boundary condition $\rho_{-n} = 0 $ for $n \geq 1$. At steady-state $\dot{\rho}_m = 0$, this recursive relation is solved via the Frobenius method. Let $A(z) = \sum_m \rho_m z^m$. We multiply both sides by $z^m$ and summing over $m$. This gives
\begin{align}
    (1-z^2)\gamma_2 A''(z) + (1-z)(\kappa-\gamma_1 z) A'(z) - (1-z) \gamma_1 A(z) &= 0 \\
    (1+z)\gamma_2 A''(z) + (\kappa-\gamma_1 z) A'(z) - \gamma_1 A(z) &= 0  
\end{align}
Let $a = \kappa/\gamma_2, b=\gamma_1/\gamma_2$. Furthermore, define $u = (1+z)b$, the above second-order ODE takes the form of confluent-hypergeometric equations
\begin{align}
    uA''(u) + (a+b - u)A'(u) - A(u) = 0
\end{align}
whose solution regular at $u=0$ is
\begin{align}
    A(u) = c \; {}_1 F_1(1; a+b; u) = c \; {}_1 F_1 \left(1; \frac{\kappa}{\gamma_2} + \frac{\gamma_1}{\gamma_2}; \frac{\gamma_1}{\gamma_2} (1+z) \right)
\end{align}
Setting $z = 1$ and using the fact that  $\sum_{m}\rho_m = 1$, we find the normalization coefficient to be
\begin{align}
    c = \frac{1}{{}_1 F_1 \left(1; \frac{\kappa}{\gamma_2} + \frac{\gamma_1}{\gamma_2}; 2\frac{\gamma_1}{\gamma_2} \right)}
\end{align}
To extract $\rho_m$, we use the fact that
\begin{align}
    \rho_m &= \frac{1}{m!}\frac{d^m}{dz^m} A(z)|_{z=0} \\
    &= c \frac{1}{m!} \frac{d^m}{dz^m} {}_1 F_1 \left(1; a+b; b (1+z) \right)|_{z=0} \\
    &= c \frac{1}{m!} b^m \frac{d^m}{du^m} {}_1 F_1 \left(1; a+b; u \right)|_{u=b} \\
    &= c \frac{b^m}{(a+b)_m} {}_1 F_1(1+m; a+b+m; b)
\end{align}
Writing in terms of the original variable 
\begin{align}
    \rho_m = \frac{(\gamma_1/\gamma_2)^m}{((\kappa+\gamma_1)/\gamma_2)_m} \frac{{}_1 F_1(1+m; (\kappa+\gamma_1)/\gamma_2+m; \gamma_1/\gamma_2)}{{}_1 F_1(1; (\kappa+\gamma_1)/\gamma_2; 2\gamma_1/\gamma_2)}, \qquad \rho = \sum_{m=0}^\infty \rho_m \ket{m}\bra{m}
\end{align}
gives the steady-state density matrix. 

We can use the steady-state density matrix to compute observables. The mean photon number is given by
\begin{align}
    \langle \hat{n} \rangle = \sum_{m=0}^\infty m \frac{(\gamma_1/\gamma_2)^m}{((\kappa+\gamma_1)/\gamma_2)_m} \frac{{}_1 F_1(1+m; (\kappa+\gamma_1)/\gamma_2+m; \gamma_1/\gamma_2)}{{}_1 F_1(1; (\kappa+\gamma_1)/\gamma_2; 2\gamma_1/\gamma_2)}
\end{align}
To facilitate calculation, we define generating function $Z(2\alpha) = {}_1 F_1(1; \beta, 2\alpha)$ with $\beta = (\kappa + \gamma_1)/\gamma_2, \alpha = \gamma_1/\gamma_2$. Then, $\rho_m = \omega_m/Z$ with $\omega_m(\alpha) = \frac{\alpha^m}{(\beta)_m}{}_1 F_1(1+m; \beta+m; \alpha)$. We note that
\begin{align}
    \sum_{m=0}^\infty \alpha \frac{d}{d\alpha} \omega_m(\alpha) = 2 \sum_{m=0}^\infty m \omega_m(\alpha) \implies \langle \hat{n} \rangle = \alpha \frac{\partial}{\partial (2\alpha)}\ln(Z(2\alpha))
\end{align}
where we used $a \frac{d}{d\alpha}\omega_m(\alpha) = m \omega_m(\alpha) + (m+1)\omega_{m+1}(\alpha)$.
The mean photon number is
\begin{align}
    \langle \hat{n} \rangle = \frac{1}{Z(2\alpha)} \frac{\alpha}{\beta} {}_1 F_1(2; \beta+1; 2\alpha) = \frac{\alpha}{\beta} \frac{{}_1 F_1(2; \beta+1; 2\alpha)}{{}_1 F_1(1; \beta; 2\alpha)}
\end{align}
When $\kappa = 0$, the semiclassical EOM of the vdP oscillator is $\dot{\alpha} = (\gamma_1/2 - \gamma_2 |\alpha|^2)\alpha $, from which we find the semiclassical mean photon number to be $n_{\rm semi} = \gamma_1 / 2\gamma_2$. Indeed, we find
\begin{align}
    \langle \hat{n} \rangle = \frac{{}_1 F_1(2; \gamma_1/\gamma_2+1; 2\gamma_1/\gamma_2)}{{}_1 F_1(1; \gamma_1/\gamma_2; 2\gamma_1/\gamma_2)}  \qquad \xrightarrow{\gamma_1/\gamma_2 \rightarrow \infty} \qquad  \frac{\gamma_1}{2\gamma_2}
\end{align}

The generating function method becomes especially useful when we compute higher-moment correlation function, since it allows us to relate the correlation function to derivatives of hypergeometric functions. Define $D = \alpha \frac{d}{d(2\alpha)}$. Using $Z(2\alpha) = {}_1 F_{1}(1; \beta; 2\alpha)$, we have
\begin{align}
    \langle \hat{n} \rangle = D \ln Z(2\alpha) = \alpha \frac{Z'(2\alpha)}{Z(2\alpha)}
\end{align}
A direct computation shows that
\begin{align}
    \frac{1}{Z}\sum_{m=0}^\infty D^2 \omega_m(\alpha) = \frac{1}{Z}\sum_{m=0}^\infty\left( \frac{\alpha}{2}\frac{d}{d(2\alpha)} \omega_m(\alpha) + \alpha^2 \frac{d^2}{d(2\alpha)^2} \omega_m(\alpha) \right)= \langle \hat{n}^2 \rangle - \frac{1}{2} \langle \hat{n} \rangle  
\end{align}
from which we find the second moment to be
\begin{align}
    \langle \hat{n}^2 \rangle = \alpha^2\frac{Z''(2\alpha)}{Z(2\alpha)} + \alpha \frac{Z'(2\alpha)}{Z(2\alpha)}
\end{align}
The variance is given by
\begin{align}
    \langle \hat{n}^2 \rangle - \langle \hat{n} \rangle^2 &= \alpha^2\frac{Z''(2\alpha)}{Z(2\alpha)} + \alpha \frac{Z'(2\alpha)}{Z(2\alpha)} - \alpha^2 \left( \frac{Z'(2\alpha)}{Z(2\alpha)}\right)^2 \\
    &= \frac{\alpha}{Z \beta} \left( \frac{2\alpha}{\beta+1} {}_1F_1(3; \beta+2 ; 2\alpha) + {}_1 F_1(2; \beta+1; 2\alpha) - \frac{\alpha}{Z \beta}({}_1 F_1(2;\beta+1; 2\alpha)^2) \right)
\end{align}
where $Z' = \frac{1}{\beta}{}_1 F_{1}(2;\beta+1; 2\alpha), Z'' = \frac{2}{\beta(\beta+1)} {}_1 F_{1}(3; \beta+2; 2\alpha) $. The Fano ratio is given by
\begin{align}
    F = \frac{\langle n^2 \rangle - \langle n \rangle^2}{\langle n \rangle} = 1 + \frac{\alpha}{\beta}\left( \frac{2\beta}{\beta+1} \frac{{}_1 F_1(3; \beta+2; 2\alpha)}{{}_1 F_1(2; \beta+1; 2\alpha)} - \frac{{}_1 F_1(2; \beta+1; 2\alpha)}{{}_1 F_1(1; \beta; 2\alpha)} \right) 
\end{align}
When $\kappa = 0$, the large photon limit ($\gamma_1/\gamma_2 \rightarrow \infty$) gives $F = 1.5$. This shows that the Fock space distribution is not Poissonian. Furthermore, the Fano ratio increases as we increase $\kappa$.

\end{document}